\documentclass[showpacs,aps,pra,twocolumn,10pt,superscriptaddress]{revtex4-2}
\usepackage{amsmath,graphicx,color}
\usepackage{comment}

\usepackage[bookmarks=false]{hyperref}
\usepackage[normalem]{ulem} %necessary to use \sout

\begin{document}

\title{Optomechanically  induced transparency in the presence of a strong Duffing nonlinearity}

\author{Rong Zhang}
\affiliation{Beijing Computational Science Research Center, Beijing 100193, China}

\author{Jing Qiu}
\affiliation{Beijing Computational Science Research Center, Beijing 100193, China}
%\affiliation{Southwest Institute of Technical Physics, Chengdu, Sichuan 610041, China}

\author{Stefano Chesi}
\email[]{stefano.chesi@csrc.ac.cn}
\affiliation{Beijing Computational Science Research Center, Beijing 100193, China}
\affiliation{Department of Physics, Beijing Normal University, Beijing 100875, China}

\author{Yingdan Wang}
\email[]{yingdan.wang@itp.ac.cn}
\affiliation{Institute of Theoretical Physics, Chinese Academy of Sciences, Beijing 100190, China}
\affiliation{School of Physical Sciences, University of Chinese Academy of Sciences, Beijing 100049, China}

\date{\today}
\begin{abstract}
We consider the combined effect of a Duffing nonlinearity and the optomechanical interaction on the cavity density of states (DOS),  which is directly observable in optomechanically induced transparency (OMIT) experiments. Both nonlinearities introduce the same type of interaction between polaritons, producing a distinct feature in the optical response. We derive the resonant condition that enhances the nonlinear scattering between polaritons and study the parameter dependence of the cavity DOS in detail. This allows us to characterize the typical strength and optimal conditions for observing the quantum effects of the Duffing nonlinearity in OMIT experiments.
\end{abstract}


\maketitle

\section{\label{sec:level1}Introduction}

Optomechanical systems exploit interactions between electromagnetic (optical or microwave) modes and mechanical oscillators~\citep{arevopto}, and have promising applications to quantum technologies such as quantum sensing and transduction~\cite{2022NatPhys_review}. Currently, however, most optomechanical devices rely on a parametrically-enhanced linear interaction, whose effects largely dominate the nonlinear character of the bare radiation pressure term. This linear interaction has been pushed to the strong~\cite{normalmode1,normalmode2} and ultra-strong~\cite{2019PRL_Teufel} coupling regimes. It has also enabled sideband cooling to the mechanical ground state~\cite{coolex2,coolex3}, as well as the generation of squeezed~\cite{2013Nature_Painter,2015Science_Schwab} and entangled~\cite{2018Nature_Sillanpa,2018Nature_Riedlinger,2021Science_Teufel} states.

In principle, the nonlinear character of the bare optomechanical interaction could allow for a wider variety of phenomena and protocols, such as generating cat states~\cite{tombesi1997,cat1,nongaussian2}, non-Gaussian~\citep{qucl1,Girvin2011,darkstate2013} and non-classical~\citep{nonclassical12012,nonclassical22013,Lijie2013} mechanical states, the photon blockade~\citep{rabl2011}, and quantum non-demolition detection of phonon number~\citep{phononback3}. In most of these instances, however, the weakness of the optomechanical coupling represents a major obstacle.  Although these limitations could be alleviated in multi-mode setups~\citep{twomodef2012,twomoder2012,2015_PRA_Xu_Taylor,collective2012, lijing2018,qiujing2022}, it is still desirable to maintain the simplicity of a single optomechanical cavity. In such a system, early theoretical studies~\cite{nonlinear1,florian2013,teuful2013} have proposed optomechanically induced transparency (OMIT) as a direct and sensitive probe of quantum effects induced by the nonlinearity, but these OMIT signatures remain elusive. 

In this article, we propose a strategy to circumvent this problem by taking advantage of a sufficiently strong Duffing term. This type of nonlinearity has been extensively studied in nanomechanical~\cite{duffingex1,duffingex3,roukes2013} and optomechanical~\cite{duffingex2,duffingex5,nori2015,nonlinear2015,duffingex4,dissipative2016} systems. Although the Duffing term is purely mechanical, it can affect the response of the cavity mode through the optomechanical interaction. In particular, it has been shown how the nonlinear oscillations and bistability of the mechanical mode can be mapped to the OMIT signal~\cite{duffingex2,duffingex4,nonlinear2015,duffingex5}. This, however, requires a relatively strong probe field, outside the linear-response regime of interest here. Furthermore, the OMIT transmission is determined in those experiments by the classical dynamics of the Duffing oscillator~\cite{nonlinear2015}, thus does not reflect genuine quantum effects. Quantum state engineering in a driven optomechanical cavity with a strong Duffing term has been explored in Ref.~\cite{nori2015}. That scheme, while leading to strong mechanical squeezing, is still based on linearized interactions, which can only access Gaussian states.

To identify observable quantum-mechanical signatures of the nonlinearity, we analyze here the characteristic dip in the OMIT signal predicted in Refs.~\cite{nonlinear1,florian2013,teuful2013}, extending those analyses to include a strong Duffing term. The characteristic dip occurs with a probe frequency $\omega_{pr}$ detuned from the pump frequency $\omega_{l}$ by approximately \emph{twice} the mechanical frequency $\Omega_m$~\cite{nonlinear1,florian2013,teuful2013}. This condition differs from standard OMIT (where $\omega_{pr}-\omega_l \simeq \Omega_m$) and reflects the presence of a significant three-wave mixing interaction $\propto \hat{c}^\dag_-\hat{c}^\dag_-\hat{c}_+ $, where $\hat{c}_+ $ is a photon-like polariton and $\hat{c}_- \simeq \hat{b}$ is predominantly mechanical. Our main conclusion is that the OMIT signatures associated with this effective interaction can be significantly enhanced by a strong Duffing nonlinearity.

More quantitatively, the size of the dip is controlled by an effective cooperativity $C_{eff}$ which, in the absence of Duffing nonlinearity, is approximately given by~\cite{nonlinear1}:
\begin{equation}\label{Ceff_ref_noDuffin}
C_{eff}\sim\frac{g^2}{\kappa^{2}},
\end{equation}
where $g$ is the bare optomechanical coupling and $\kappa$ the cavity linewidth. Since the largest realized values of the `quantum parameter'~\cite{qucl1} are of order $g/\kappa \sim 0.01$~\cite{normalmode2,2012_APL_Painter,nongaussian2}, nonlinear features in OMIT experiments are extremely challenging to observe. Here, however, we find that the same type of OMIT signatures can be obtained in the presence of a Duffing term. The mechanical nonlinearity becomes the dominant contribution to $C_{eff}$ when $\eta \gg g^2/\Omega_m$. In this regime, the effective cooperativity is given by:
\begin{equation}\label{Ceff_estimate_Duffing}
C_{eff}\sim\frac{\eta \Omega_m}{\kappa^{2}},
\end{equation} 
which is independent of $g$ and much larger than $g^2/\kappa^2$ [see Eq.~(\ref{Ceff_ref_noDuffin})], facilitating experimental detection.

Our work has interesting implications beyond OMIT. First, as the enhancement of $C_{eff}$ is directly related to the presence of a stronger three-wave mixing interaction, we expect that the system will also facilitate the generation of non-classical states. For example, a recent proposal to implement mechanical cat states~\cite{nongaussian2} is based on an analogous two-tone driving scheme and exploits the same type of effective nonlinear interaction that we analyze here in relation to OMIT. 

Furthermore, the mechanism we exploit should be applicable to alternative schemes where a different type of nonlinear term dominates, such as a significant Kerr interaction~\cite{2021_PRR_Steele,2023_NatComm_Steele,Cai_2025}. Indeed, in our setup the two types of nonlinear interactions play conceptually distinct roles: the optomechanical coupling mainly acts via its linearized contribution, parametrically enhanced by the external drive, to hybridize the optical and mechanical modes. Following this hybridization, a sufficiently strong Duffing term naturally leads to the desired nonlinear interaction between photon- and phonon-like polaritons. Therefore, the specific form of the dominant nonlinearity (here the Duffing term) should not play a crucial role for the effects we characterize.

The outline of our paper is as follows: We begin in Sec.~\ref{sec:SYSTEM-MODEL} by introducing the model. We then express in Sec.~\ref{sec:SYSTEMS-IN-POLARITON} the optomechanical system in terms of polaritons, and discuss the parameters realizing the resonant condition. In Sec.~\ref{sec:Effect-of-Nonlinear} we analyze in detail the signatures of the optomechanical and Duffing nonlinearities on the cavity density of states, directly related to OMIT. Finally, our conclusions are given in Sec.~\ref{sec:Conclusion}.

%%%%%%%%%%%%%%%%%%%%%%%%%%%%%%%%%%%%%%%
\begin{figure}
\begin{centering}
\includegraphics[scale=0.29]{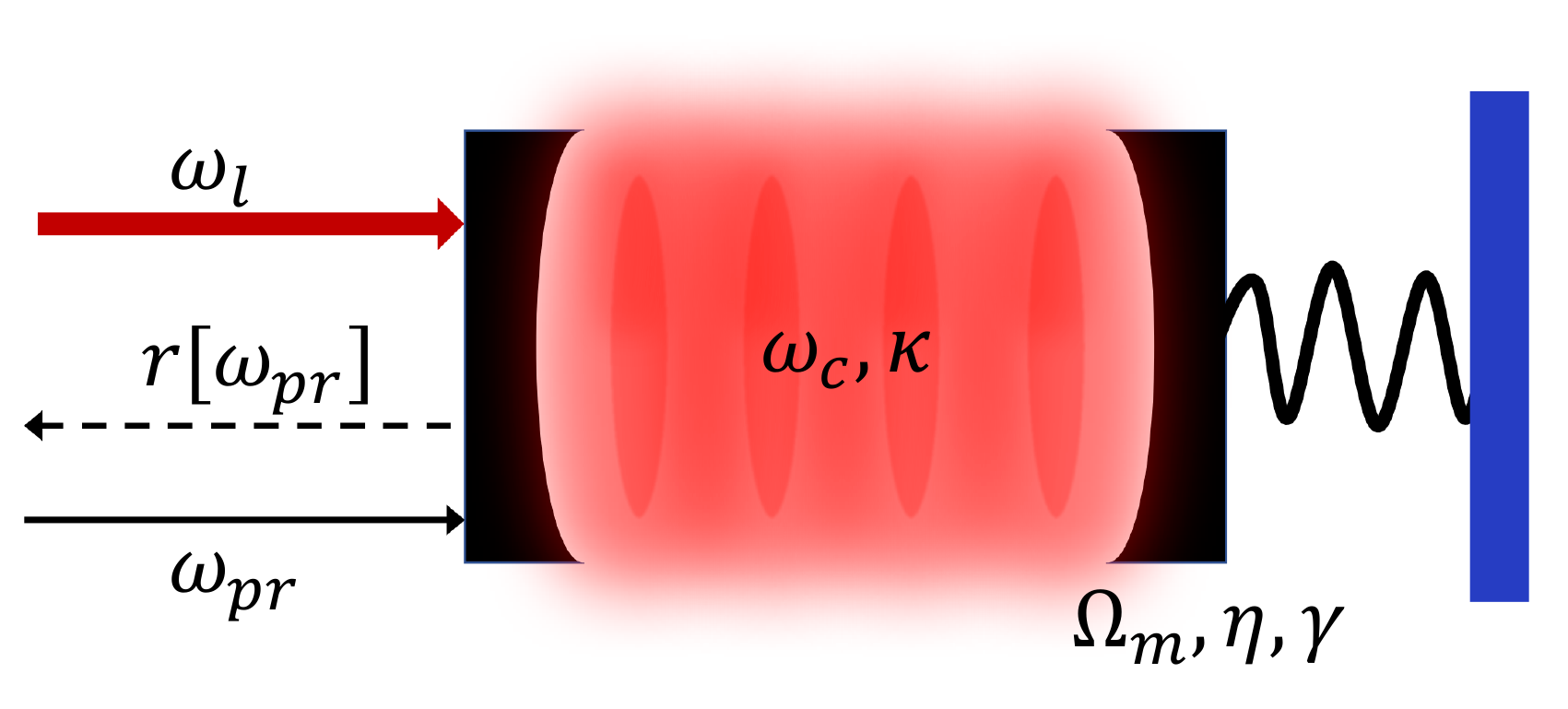}
\par\end{centering}
\caption{Schematic illustration of an OMIT setup, where $\omega_{c}$ is the cavity frequency, $\kappa$ the cavity linewidth, $\Omega_{m}$ the mechanical frequency, $\gamma$ the mechanical damping rate, and $\eta$ the Duffing coefficient.
The red arrow represents a strong laser drive of frequency $\omega_{l}$, while $\omega_{pr}$ is the frequency of a probe field, with corresponding reflection coefficient $r\left[\omega_{pr}\right]$.\label{fig:setup}}
\end{figure}
%%%%%%%%%%%%%%%%%%%%%%%%%%%%%%%%%%%%%%%

\section{\label{sec:SYSTEM-MODEL} MODEL}

We consider an optomechanical system (shown in Fig. \ref{fig:setup}),
in which the mechanical mode has a significant Duffing nonlinearity.
The system is described by the Hamiltonian $\hat{H}=\hat{H}_{0}+\hat{H}_{dr}+\hat{H}_{diss}$,
with:
\begin{equation}\label{H0}
\hat{H_{0}}=\omega_{c}\hat{a}^{\dagger}\hat{a}-g\left(\hat{b}+\hat{b}^{\dagger}\right)\hat{a}^{\dagger}\hat{a}+\Omega_{m}\hat{b}^{\dagger}\hat{b}+\frac{\eta}{4}(\hat{b}^{\dagger}+\hat{b})^{4}.
\end{equation}
Here, $\hat{a}$ is the annihilation operator of the cavity mode, $\omega_{c}$
is the cavity frequency, $\hat{b}$ is the annihilation operator for the mechanical
mode, $\Omega_{m}$ is the mechanical frequency, and $g$ is optomechanical single-photon
coupling strength. The Duffing coefficient $\eta$ could be positive or negative, which respectively leads to stiffening or softening of the oscillator~\cite{duffingbook}. In the following discussion we assume
a positive Duffing coefficient, as is almost always the case in optomechanical systems \citep{duffingex1,duffingex2,duffingex3,duffingex4,duffingex5,roukes2013}.
An artificial positive Duffing term would also be generated by coupling mechanics to a qubit, which is an effective approach
to engineer large values of $\eta$~\cite{nonge1,nori2015}. For a negative Duffing term, we expect similar conclusions. $\hat{H}_{dr}$
is the drive term:
\begin{equation}\label{Hdr}
\hat{H}_{dr}=\sqrt{\kappa}\bar{\alpha}_{in}e^{i(\omega_{c}-\Delta)t}\hat{a}+\mathrm{H.c.},
\end{equation}
where we will only consider a pump with a positive value of $\Delta = \omega_{c}-\omega_{l}$. Finally,
 $\hat{H}_{diss}$ models Markovian environments for the cavity and mechanical modes, with damping rates $\kappa$ and $\gamma$, respectively. The detailed form of $\hat{H}_{diss}$ is given in Appendix~\ref{sec:bath}.

To analyze the system, we first transform the cavity mode to the rotating frame at the laser frequency
$\omega_{l}$, and perform displacement transformations $\hat{a}\rightarrow\alpha+\hat{a}$ and
$\hat{b}\rightarrow\beta+\hat{b}$ where the classical amplitudes in the steady state are obtained by solving:
\begin{align}
&-i\left(\Delta-2g\beta\right)\alpha-\frac{\kappa}{2}\alpha-i\sqrt{\kappa}\bar{\alpha}_{in}=0,\label{eq:steady1} \\
&\Omega_{m}\beta+8\eta\beta^3-g|\alpha|^{2}=0.\label{eq:steady2}
\end{align}
 In contrast to previous OMIT experiments with significant Duffing nonlinearity~\cite{duffingex2,duffingex4}, we can neglect the effect of the probe field on the classical amplitudes. This is because in this work we will only analyze the linear response to a sufficiently weak probe. While OMIT can be used to map the well-known nonlinear oscillations and bistability of the classical Duffing oscillator~\cite{duffingex2,duffingex4}, observing those effects requires stronger probe fields, beyond the linear response regime. We also note that the standard regime to observe OMIT requires the pump and probe fields to satisfy $\omega_{pr} - \omega_l \simeq \Omega_m$, which induces a beat frequency resonant with the mechanical element~\cite{duffingex2,duffingex4}. In our work, however, the preferential regime to observe signatures of the nonlinear interaction is with $\omega_{pr} - \omega_l > 2\Omega_m$, thus the beat frequency is far from the mechanical resonance. Under these conditions, the system remains close to the classical steady-state amplitudes of Eqs.~(\ref{eq:steady1}) and (\ref{eq:steady2}), determined by the radiation pressure of the strong pump field.

The above transformations result in an Hamiltonian $\hat{H}_d$, given explicitly by Eq.~(\ref{Hd}) of Appendix~\ref{sec:bath}. $\hat{H}_d$ is further simplified by applying a squeezing transformation~\citep{nori2015}, $\hat{S}(r)=\exp[r(\hat{b}^{2}-\hat{b}^{\dagger2})/2]$, where
\begin{equation}\label{r_squeezing}
r=\frac14 \ln\left(1+\frac{24\eta\beta^{2}}{\Omega_{m}}\right).
\end{equation}
 The final Hamiltonian $\hat{H}_s = \hat{S}(r)^\dag \hat{H}_d \hat{S}(r)$ takes  the form $\hat{H}_s=\hat{H}_{l}+\hat{H}_{nl}+\hat{H}_{s,diss}$, where $\hat{H}_{l}$ describes the quantum linear interaction between cavity and mechanics and $\hat{H}_{nl}$ includes the intrinsic optomechanical and Duffing quantum nonlinearities. Explicitly, $\hat{H}_{l}$ reads:
\begin{equation}\label{eq:hamiltonian}
\hat{H_{l}}=\Delta_{c}\hat{a}^{\dagger}\hat{a}+\widetilde{\Omega}_{m}\hat{b}^{\dagger}\hat{b}-G_{s}\left(\hat{a}^{\dagger}+\hat{a}\right)\left(\hat{b}^{\dagger}+\hat{b}\right),
\end{equation}
where $\Delta_{c}=\Delta-2g\beta$ gives the shifted detuning
and $\widetilde{\Omega}_{m}=\Omega_{m}e^{2r}$ is the squeezed and stiffened mechanical frequency. The linear optomechanical coupling $G=g\alpha$ is also affected by the squeezing of the mechanics, giving $G_{s}=Ge^{-r}$. The nonlinear terms, after performing the linearization and squeezing procedure, read:
\begin{align}\label{Hnl}
\hat{H}_{nl}\simeq &-ge^{-r}\hat{a}^{\dagger}\hat{a}\left(\hat{b}^{\dagger}+\hat{b}\right)+2\eta\beta e^{-3r}\left(\hat{b}+\hat{b}^{\dagger}\right)^{3}\nonumber \\
& +\frac14 \eta e^{-4r}\left(\hat{b}+\hat{b}^{\dagger}\right)^{4}.
\end{align}
Here, for simplicity, we use the original symbols $\hat a,\hat b$ to denote the transformed cavity and mechanical modes. Finally, as discussed in Appendix~\ref{sec:bath}, the Hamiltonian of the bath $\hat{H}_{s,diss}$ is identical to  the case without Duffing nonlinearity,  except for the rescaling:
\begin{equation}\label{gamma squeezed}
\gamma \to \gamma e^{-2r},
\end{equation}
induced by the squeezing transformation of the mechanical coordinate.

The above treatment is essentially identical to the one in Ref.~\cite{nori2015}. In particular, like in that work,  we have dropped the $\gamma$-dependent terms in Eq.~(\ref{eq:steady2}), which is appropriate in the regime of interest where $\gamma\ll \sqrt{\eta\Omega_{m}} \ll \Omega_{m}$. As a result, $\beta$ is real and we will also take $\alpha$ real in the following, which can be always achieved choosing an appropriate phase for the drive amplitude $\bar\alpha_{in}$. There is, however, a small discrepancy between Eq.~(\ref{eq:steady2}) and the corresponding Eq.~(5b) of Ref.~\cite{nori2015}, where an additional term $\propto \eta\beta$ is present. This difference is due to our choice of expressing the nonlinear interactions in terms of the $b+b^\dag$ quadrature, which transforms in a simple way under $S(r)$, see Eq.~(\ref{Hnl}). Instead, Ref.~\cite{nori2015} considers the normal form of the nonlinear interaction, and the significantly more complex expression obtained after the squeezing transformation is not given explicitly. Similarly, the squeezing parameter of Ref.~\cite{nori2015} is obtained by substituting $\beta^2 \to \beta^2 +1/4 $ in Eq.~(\ref{r_squeezing}). The two descriptions must clearly be equivalent when the nonlinear terms are accounted for. Even at the level of the linear Hamiltonian the difference is negligible, since the classical amplitudes satisfy $\alpha^2,\beta^2 \gg 1$ (thus, $\beta^2 +1/4 \simeq \beta^2$).

%%%%%%%%%%%%%%%%%%%%%%%%%%%%%%%%%%%%%%%%%%%%%%%%%%%%%%%%%%%%%%%%%%%%%%%%%%%%%%%%%%%%%%%%
\section{RESONANT POLARITON MODES}\label{sec:SYSTEMS-IN-POLARITON}
%%%%%%%%%%%%%%%%%%%%%%%%%%%%%%%%%%%%%%%%%%%%%%%%%%%%%%%%%%%%%%%%%%%%%%%%%%%%%%%%%%%%%%%%

An effective strategy to probe the effects of a weak optomechanical non-linearity is by inducing resonant three-wave mixing terms between polariton eigenmodes~\cite{nonlinear1,teuful2013,florian2013,lijing2018,qiujing2022}. The Bogoliubov transformation diagonalizing $H_l$ is of the form:
\begin{equation}\label{eq:trans}
\left(\begin{array}{cc}
\hat{a} & \hat{b}\end{array}\right)^{T}=V\left(\begin{array}{cccc}
\hat{c}_{-} & \hat{c}_{+} & \hat{c}_{-}^{\dagger} & \hat{c}_{+}^{\dagger}\end{array}\right)^{T}
\end{equation}
where $T$ denotes transposition and $V$ is given explicitly in Appendix~\ref{sec:Appendix_Bogolubov}. In terms of the $c_\pm$ modes, the Hamiltonian becomes:
\begin{align}\label{eq:polariton}
\hat{H}\simeq & E_{+}\hat{c}_{+}^{\dagger}\hat{c}_{+}+E_{-}\hat{c}_{-}^{\dagger}\hat{c}_{-} \nonumber \\
& +\left[\left(\tilde{g}_{d}+\tilde{g}_{os}\right)\hat{c}_{+}^{\dagger}\hat{c}_{-}\hat{c}_{-}+\mathrm{H.c.}\right ] +H_{diss},
\end{align}
where in the second line we only keep the nonlinear terms scattering a $+$ polariton into
two $-$ polaritons. This nonlinear term becomes resonant by enforcing the condition:
\begin{equation}\label{resonance_condition}
E_{+}=2E_{-}.
\end{equation}
 The effect of non-resonant terms (e.g.,  $\propto \hat{c}_{-}^{\dagger}\hat{c}_{+}\hat{c}_{+}$) is discussed in Appendix~\ref{app:non_resonant}, where we show that they can be safely neglected. Furthermore, in Appendix~\ref{appendix_multipolariton}, we outline the parallel treatment of the alternative resonant condition $E_{+}=3E_{-}$, for which the nonlinear interaction $\propto \hat{c}_{+}^{\dagger}\hat{c}_{-}\hat{c}_{-}\hat{c}_-$ becomes relevant, and argue that it gives rise to much smaller effects.

In Eq.~(\ref{eq:polariton}),  the expression of $E_\pm$ in terms of the parameters $\Delta_c$, $\widetilde\Omega_m$, and $G_s$ of Eq.~(\ref{eq:hamiltonian}) is completely analogous to previous works and, for convenience of the reader, is given in Eq.~(\ref{eq:enp}).  Note that a real value of $E_-$ requires:
\begin{equation}\label{stability_RH}
4 G_s^2 < \Delta_c \widetilde\Omega_m.
\end{equation}
For the range of detunings considered in this work ($\Delta_c \geq \Omega_m/2 \gg \kappa$), Eq.~(\ref{stability_RH}) coincides with the stability condition obtained more rigorously from the Routh–Hurwitz criterion~\cite{nori2015}. As we will see, when the resonance condition in Eq.~(\ref{resonance_condition}) is satisfied, the resulting parameters satisfy $G_s^2 \ll \Delta_c, \widetilde\Omega_m$, well within the stable regime of weak hybridization.

%%%%%%%%%%%%%%%%%%%%%%%%%%%%%%%%%%%%%%%%%%%%%%%%%%%%%%%%%%%%%%%%%%%%%%%%%%%%%%%%%%%%%%%%
\begin{figure*}
\begin{centering}
\includegraphics[width=0.65\textwidth]{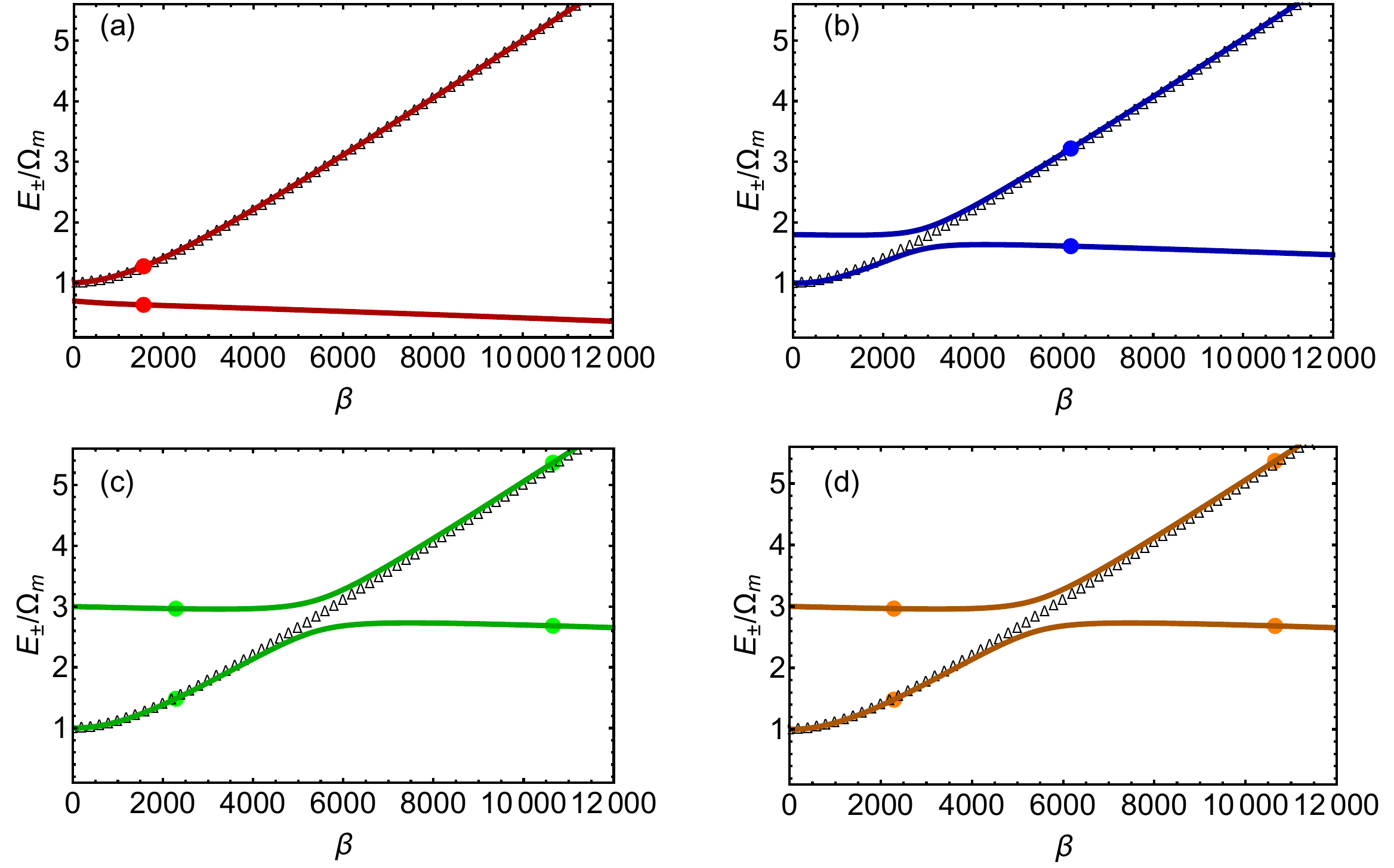}
\par\end{centering}
\caption{Dependence of the polariton energies $E_{\pm}$ on the average mechanical amplitude $\beta$. The value of $\beta$, given by Eqs.~(\ref{eq:steady1}) and~(\ref{eq:steady2}), increases monotonically with the strength of the external drive. The cavity detuning is: (a)~$\Delta=0.7\Omega_{m}$,
(b)~$\Delta=1.8\Omega_{m}$, (c)~$\Delta=2.2\Omega_{m}$, and (d)~$\Delta=3\Omega_{m}$. In each panel, the colored dots mark the values of $\beta$ realizing the resonant condition
 $E_{+}=2E_{-}$. We also show the squeezed mechanical frequency $\widetilde{\Omega}_{m}$ (triangles), to easily identify the polariton modes with a predominantly mechanical character. Other parameters are: $\kappa/\Omega_{m}=10^{-3}$,
$g/\Omega_{m}=10^{-5}$, $\eta/\Omega_{m}=10^{-8}$, $\gamma/\Omega_{m}=10^{-7}$. 
\label{fig:envsbeta}}
\end{figure*}
%%%%%%%%%%%%%%%%%%%%%%%%%%%%%%%%%%%%%%%%%%%%%%%%%%%%%%%%%%%%%%%%%%%%%%%%%%%%%%%%%%%%%%%%

In the nonlinear term of Eq.~(\ref{eq:polariton}), we distinguish between the contribution of the optomechanical interaction ($\tilde{g}_{os} \propto g$) and the Duffing term ($\tilde{g}_{d} \propto \eta$). Their expressions, obtained in a straightforward manner by applying Eq.~(\ref{eq:trans}) to $\hat{H}_{nl}$, read:
\begin{align}
&\tilde{g}_{os}= -g e^{-r}\left[ V_{11}V_{13}\left(V_{22}+V_{24}\right) \right. \nonumber \\
& \qquad\qquad\,\,\left.+\left(V_{14}V_{13}+V_{12}V_{11}\right)\left(V_{23}+V_{21}\right)\right], \label{eq:optonon} \\
&\tilde{g}_{d}=6\beta\eta e^{-3r}\left(V_{22}+V_{24}\right)\left(V_{21}+V_{23}\right)^{2}.\label{eq:duffing}
\end{align}

In general, the resonance condition Eq.~(\ref{resonance_condition}) can be enforced by choosing an appropriate drive strength. This is illustrated by the plots of $E_\pm$ in Fig.~\ref{fig:envsbeta}, where we consider various values of $\Delta$ and choose a large value of $\eta$, such that:  
\begin{equation}\label{large_eta}
\overline\eta \equiv \frac{\Omega_m\eta}{g^2}  \gg 1 .
\end{equation}
This regime is particularly interesting for us since, as we will show, the Duffing nonlinearity plays a dominant role. Instead, the opposite limit $\overline\eta \ll 1$ only introduces marginal changes to the treatment of Ref.~\cite{nonlinear1}. 

When Eq.~(\ref{large_eta}) is satisfied, the resonant condition is relatively easy to discuss, because it is always realized with weakly hybridized optical and mechanical modes. Therefore, as also seen in Fig.~\ref{fig:envsbeta}, the polariton frequencies at the resonant points (marked with dots on the polariton branches) are close to the bare values $\Delta$ and $\widetilde\Omega_m$. Since the squeezed mechanical frequency $\widetilde\Omega_m =\Omega_m e^{2r}$ is always enhanced by the drive, the resonance can only be realized for $\Delta > \Omega_m/2$. Above this value, we can distinguish between two cases:

(i) If $\Omega_m/2 < \Delta < 2\Omega_m$, we have the situation illustrated by the two upper panels of Fig.~\ref{fig:envsbeta}. Here, by increasing sufficiently the drive strength, a single resonant point can be reached. The upper polariton has $E_+ \simeq \widetilde\Omega_m$ and is predominantly phononic, while the lower polariton is at $E_- \simeq \Delta$. Setting $\widetilde\Omega_m \simeq 2\Delta$, and making use of Eq.~(\ref{r_squeezing}), we easily find:
\begin{equation}
\beta_{res}\simeq\sqrt{\left(4\Delta^{2}-\Omega_{m}^{2}\right)/\left(24\eta\Omega_{m}\right)}.\label{eq:beta_res1}
\end{equation}

(ii) If $\Delta > 2\Omega_m$, as shown in the two lower panels of Fig.~\ref{fig:envsbeta}, there are two resonant points. The one at large drive is completely analogous to the previous case, i.e., $\beta$ is given by Eq.~(\ref{eq:beta_res1}) and the upper (lower) polariton has a phononic (photonic) character. For the other resonant point, instead, the character of the polaritons is swapped, i.e., the upper one is photonic, at  $E_+ \simeq \Delta$, and the lower one is phononic, at $E_- \simeq \widetilde\Omega_m$. This resonance occurs for:
\begin{equation}
\beta_{res}\simeq\sqrt{\left(\Delta^{2}-4\Omega_{m}^{2}\right)/\left(96\Omega_{m}\eta\right)},\label{eq:beta_res2}
\end{equation}
which, as clear from panels (c) and (d) of Fig.~\ref{fig:envsbeta}, is realized at a much smaller driving strength. There are two other benefits of this resonant point. First, since the driving strength is relatively weak, the squeezing parameter $r$ does not suppress excessively the effective nonlinear couplings. In fact, we have $\tilde{g}_{os}\propto e^{-r}$ and $\tilde{g}_{d}\propto e^{-3r}$, as can be seen from Eqs.~(\ref{eq:optonon}) and (\ref{eq:duffing}), respectively. The other benefit is that the nonlinear signatures in the OMIT lineshape should be easier to detect when the upper polariton is photonic and the lower polariton is phononic. In this case,  the linear response to a weak probe field shows a sharp dip appearing on top of the broad cavity lineshape~\cite{nonlinear1,teuful2013}. This signature can be more easily recognized in experiments, even for small nonlinearity. Therefore, for $\Delta>2\Omega_{m}$ we will only focus on the lower-drive resonant point.

%%%%%%%%%%%%%%%%%%%%%%%%%%%%%%%%%%%%%%%%%%%%%%%%%%%%%%%%%%%%%%
\begin{figure*}
\begin{centering}
\includegraphics[width=0.7\textwidth]{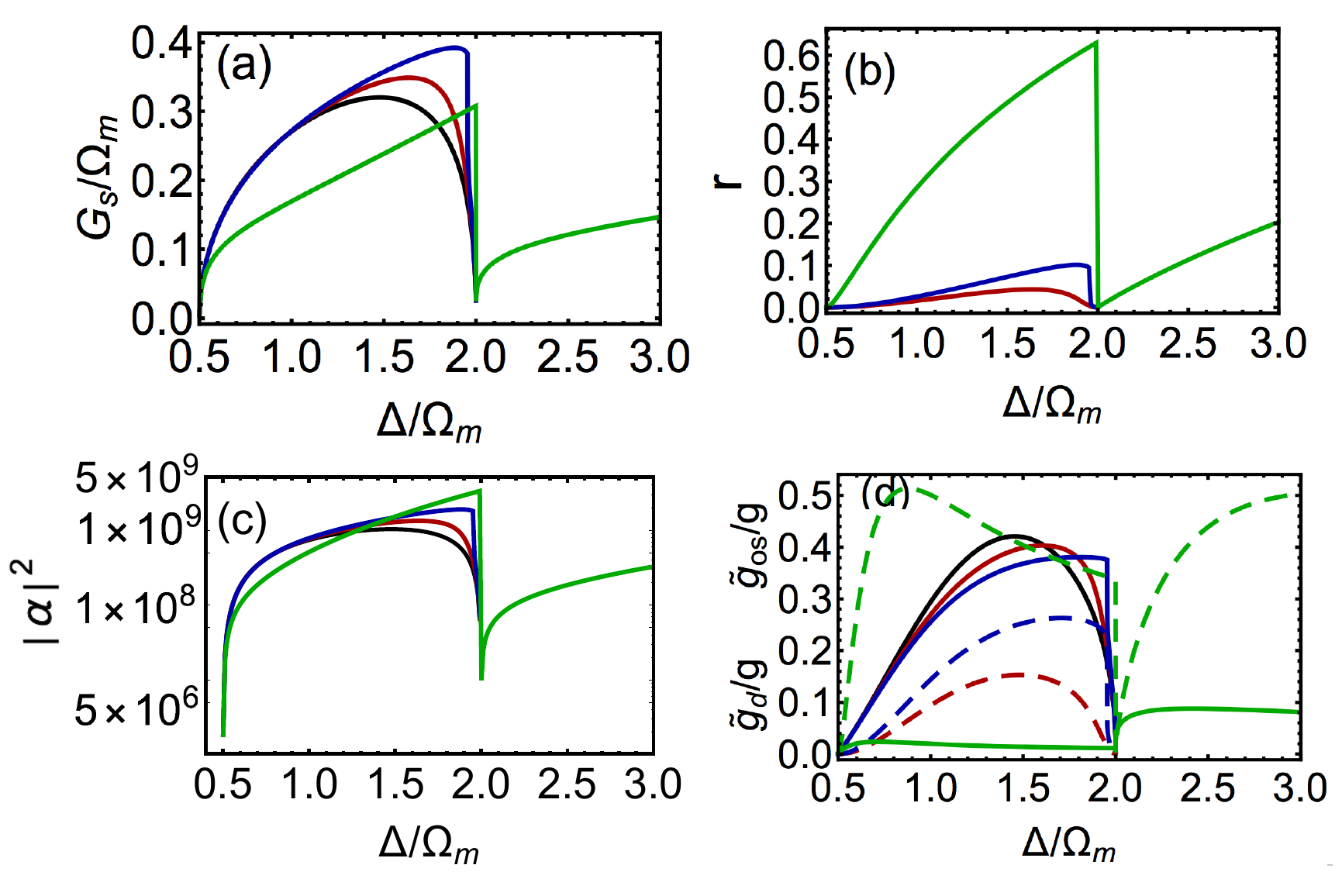}
\end{centering}
\caption{Dependence on $\Delta$ of the (a) dressed optomechanical coupling $G_s$, (b) squeezing parameter $r$, (c) number of cavity photons, and (d) effective nonlinear couplings. These quantities are evaluated at the resonant condition~(\ref{resonance_condition}), using $\eta =0$~(black), $\eta/\Omega_{m}=5\times10^{-11}$~(red), $\eta/\Omega_{m}=8\times10^{-11}$~(blue), and $\eta/\Omega_{m}=10^{-8}$~(green). In panel (d), the solid ($\tilde{g}_{os}$) and dashed ($\tilde{g}_{d}$) curves refer to the effective couplings induced by optomechanical and Duffing nonlinearities, respectively. Other parameters are: $\kappa/\Omega_{m}=10^{-3}$,
$g/\Omega_{m}=10^{-5}$, and $\gamma/\Omega_{m}=10^{-7}$. Following Eq.~(\ref{large_eta}), the rescaled values of $\eta$ are $\overline{\eta}=0,0.5,0.8,100$.}\label{fig:input-1}
\end{figure*}
%%%%%%%%%%%%%%%%%%%%%%%%%%%%%%%%%%%%%%%%%%%%%%%%%%%%%%%%%%%%%%%%

By imposing the resonant condition, we can obtain as a function of $\Delta$ several interesting quantities, plotted in Fig.~\ref{fig:input-1}. The red and blue curves refer to intermediate values of $\eta$ (i.e., yielding $\overline\eta \sim 1$) and are restricted to the interval $\Omega_m/2 < \Delta < 2\Omega_m$. This is because, as in the case without Duffing nonlinearity, the resonant condition cannot be reached within the stable regime when $\Delta>2\Omega_m$~\cite{nonlinear1}. Still, the Duffing non-linearity is sufficiently strong to induce significant changes with respect to the $\eta=0$ case (black curves). The increase in the values of the dressed coupling $G_s$ and the intracavity photon number $|\alpha|^2$, especially in the region $\Delta \lesssim 2\Omega_m$, reflect the fact that a stronger drive needs to be applied to realize the resonant condition. At the same time, the finite value of $\eta$ results in a significant contribution to the non-linear interaction between polaritons. The blue and red solid curves of Fig.~\ref{fig:input-1}(d) show that the values of $\tilde{g}_{os}$ are only weakly affected by $\eta$. Instead, the non-linear coupling $\tilde{g}_d$ due to the Duffing non-linearity (red and blue dashed curves) starts to play an important role, reaching values comparable to $\tilde{g}_{os}$.

In the regime of large Duffing nonlinearity of Eq.(~\ref{large_eta}), the behavior is qualitatively different. As shown by the green curves of Fig.~\ref{fig:input-1} (where $\overline\eta = 100$), an additional resonant branch is now allowed at $\Delta > 2\Omega_m$. As noted before, this resonant condition occurs at a relatively small strength of the drive, which is reflected by the sudden drop in $G_s$ and $|\alpha|^2$. At the same time, the squeezing parameter $r$ also drops to small values. The nonlinear interaction between polaritons is analyzed in  Fig.~\ref{fig:input-1}(d) where, as expected, $|\tilde{g}_{os}|\ll \tilde{g}_d$. The value of $\tilde{g}_d$ is generally significant in both detuning regimes ($\Omega_m/2< \Delta < 2\Omega_m$ and $\Delta > 2\Omega_m$), and with this choice of parameters can reach maximum values of approximately half the bare optomechanical coupling strength, $\max[\tilde{g}_d] \sim 0.5 g$. On the other hand, the detailed analysis in the following Section will show that the a detuning $\Delta > 2\Omega_m$ is generally more favorable (see Fig.~\ref{fig:ceff1} and Fig.~\ref{fig:ceff2}).

\begin{figure}
\begin{centering}
\includegraphics[width=0.37\textwidth]{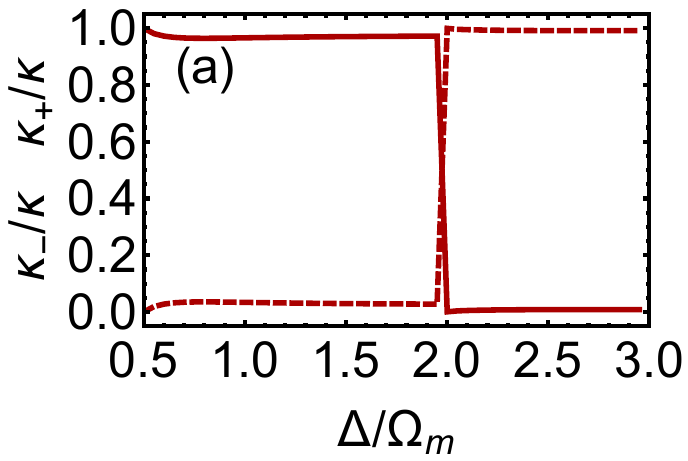}
\includegraphics[width=0.37\textwidth]{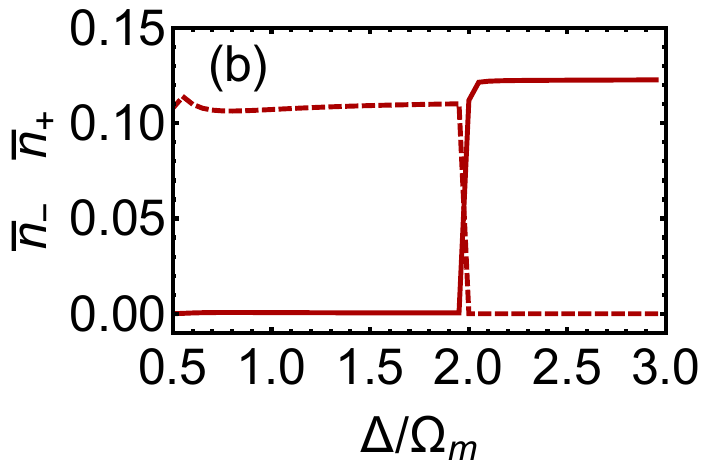}
\par\end{centering}
\caption{Effective (a) damping rates $\kappa_\pm$ and (b) occupancies $n_\pm$ of the polariton modes. The dependence on $\Delta$ is obtained after imposing Eq.~(\ref{resonance_condition}). Solid and dashed curves refer to the lower ($E_-$) and upper ($E_+$) polaritons, respectively. We used the following parameters: $\kappa/\Omega_{m}=10^{-3}$,
$g/\Omega_{m}=10^{-5}$, $\gamma/\Omega_{m}=10^{-7}$, $k_B T=0$,  and 
$\eta/\Omega_{m}=10^{-8}$.\label{fig:input-1-1}}
\end{figure}

We finally consider the coupling of the optical and mechanical baths. In the weak dissipation limit, i.e. $\kappa,\gamma\ll E_{+},E_{-},\left|E_{+}-E_{-}\right|$, we can derive the Heisenberg-Langevin equations of the polariton modes following Refs.~\cite{nonlinear1,Lemode_PRA_2015}. We obtain the effective damping rates:
\begin{align}
&\kappa_{-}=\gamma e^{-2r}\left(V_{21}+V_{23}\right)^{2}+\kappa\left(V_{11}^{2}-V_{13}^{2}\right),\label{eq:kn} \\
&\kappa_{+}=\gamma e^{-2r}\left(V_{22}+V_{24}\right)^{2}+\kappa\left(V_{12}^{2}-V_{14}^{2}\right),\label{eq:kp}
\end{align}
and occupancies:
\begin{align}
\bar{n}_{-}=\frac{1}{\kappa_{-}}\left(\gamma e^{-2r}\left(V_{21}+V_{23}\right)^{2}\bar{n}_{\mathrm{th}}\left[E_{-}\right]+\kappa V_{13}^{2}\right),\label{eq:nn}\\
\bar{n}_{+}=\frac{1}{\kappa_{+}}\left(\gamma e^{-2r}\left(V_{22}+V_{24}\right)^{2}\bar{n}_{\mathrm{th}}\left[E_{+}\right]+\kappa V_{14}^{2}\right),\label{eq:nn-1}
\end{align}
where $\bar{n}_{\mathrm{th}}[\omega]=\left[e^{\omega/\left(k_{B}T\right)}-1\right]^{-1}$ is the Bose-Einstein
distribution function at the mechanical
bath temperature $T$. Due to non-equilibrium conditions, the effective temperatures of the polariton modes, $T_\pm = E_\pm/(k_{B}\ln[1/\bar{n}_\pm+1])$, are generally different from $T$. Mathematical steps and physical considerations leading to Eqs.~(\ref{eq:kn}--\ref{eq:nn-1}) are the same of Refs.~\cite{nonlinear1,Lemode_PRA_2015} (later applied to multi-mode setups in Refs.~\cite{lijing2018,qiujing2022}), and we do not repeat them here. The only difference of these expressions from Refs.~\cite{nonlinear1,Lemode_PRA_2015} is due to the squeezing transformation, which rescales $\gamma$ as in Eq.~(\ref{gamma squeezed}). In Fig.~\ref{fig:input-1-1} we show the characteristic dependence on $\Delta$ of these quantities, evaluated at the resonance condition in the regime of $T=0$ and $\overline\eta \gg 1$. The optical character of one of the polaritons, with an effective damping similar to the value $\kappa$ of the bare cavity, is obvious from panel (a). The other polariton is predominantly phononic and has a much smaller damping rate, dominated by the weak hybridization with the cavity mode. As seen in panel (b), the polariton with mechanical character has a larger effective occupation, close to $1/8=0.125$, while the occupation of the photonic polariton is almost negligible. These plots show a relatively weak dependence of $\kappa_\pm$ and $n_\pm$ on $\Delta$, except for the abrupt switch in the photonic/phononic character of the upper/lower polaritons at $\Delta= 2\Omega_m$.

%%%%%%%%%%%%%%%%%%%%%%%%%%%%%%%%%%%%%%%%%%%%%%%%%%%%%%%%%%%%%%%%%
%%%%%%%%%%%%%%%%%%%%%%%%%%%%%%%%%%%%%%%%%%%%%%%%%%%%%%%%%%%%%%%%

In closing this section, it is worthwhile to comment on why the occupation number of the mechanical polariton remains finite in Fig.~\ref{fig:input-1-1}(b), despite the zero-temperature limit. As can be inferred from Eq.~(\ref{eq:nn}), this effect is due to the interaction with the optical bath, induced by the finite hybridization with the cavity ($V_{13} \neq 0$). In physical terms, the polariton mode can absorb a photon from the external drive (at frequency $\omega_l$), while simultaneously exciting the optical bath (at frequency $\omega_l-E_-$). This type of process, also denoted ``quantum activation,'' is discussed in detail in Refs.~\cite{nonlinear1,Lemode_PRA_2015}. In our case, by considering $T=0$ and the conditions $\overline{\eta},\kappa/\gamma\gg1 $, we find $\bar{n}_+ \simeq 1/8$ for $\Omega_m/2<\Delta<2\Omega_m$ (and $\bar{n}_- \simeq 1/8$ for $\Delta>2\Omega_m$). A similar argument applies to the occupation of the photonic polariton which, strictly speaking, remains finite at $T= 0$ as well. In fact, we can also note that $V_{14}\neq 0$ in Eq.~(\ref{eq:nn-1}). However, the structure of the Bogoliubov transformation results in an occupation number of order $1/\overline{\eta}$ in this case, which we can safely neglect in the limit of a large Duffing nonlinearity.

%%%%%%%%%%%%%%%%%%%%%%%%%%%%%%%%%%%%%%%%%%%%%%%%%%%%%%%%%%%%%%%
\section{\label{sec:Effect-of-Nonlinear}Effect of Nonlinear Interactions on the Cavity DOS}
%%%%%%%%%%%%%%%%%%%%%%%%%%%%%%%%%%%%%%%%%%%%%%%%%%%%%%%%%%%%%%%

We will now discuss the effect of the optomechanical and Duffing nonlinearities on the optical response of the cavity mode. In particular, we will analyze the cavity density of states $\rho_a[\omega]$, which is directly accessible through a weak probe of frequency $\omega_{pr}$. The probe reflectivity is given by $R[\omega_{pr}] = 1 - 2\pi \kappa_{p} \rho_a[\omega_{pr}]$~\cite{nonlinear1,Lemode_PRA_2015,lijing2018,qiujing2022}, where $\kappa_{p}$ is the additional cavity damping induced by the input/output port of the probe field, under the assumption $\kappa_{p} \ll \kappa$~\cite{lijing2018}. Since the formal description of the system in terms of $\hat c_{\pm}$ operators is the same as in the absence of the Duffing nonlinearity, we can compute $\rho_a[\omega]$ by relying on the perturbative calculation of Ref.~\cite{nonlinear1}. The polariton retarded Green's functions is given by:
\begin{align}
G^{R}_\sigma \left[\omega\right] & =  -i\int_{-\infty}^{\infty}dt\theta\left(t\right)\left\langle \left[\hat{c}_\sigma\left(t\right),\hat{c}_\sigma^{\dagger}\left(0\right)\right]\right\rangle e^{i\omega t} \nonumber\\
&=\frac{1}{\omega-E_{\sigma}+i\kappa_{\sigma}/2-\Sigma_{\sigma}^{R}\left[\omega\right]}
\end{align}
where the second line expresses $G^{R}_\sigma[\omega]$ in terms of the exact self-energies $\Sigma_{\sigma}^{R}[\omega]$. Evaluated to lowest order in the nonlinear coupling, they are obtained as follows:
\begin{align}
&\Sigma_{-}^{R}\left[\omega\right]=\frac{4\left(\tilde{g}_{d}+\tilde{g}_{os}\right)^{2}\left(\bar{n}_{-}-\bar{n}_{+}\right)}{\omega-\left(E_{+}-E_{-}\right)+i\left(\kappa_{-}+\kappa_{+}\right)/2},\label{eq:self1}\\ 
&\Sigma_{+}^{R}\left[\omega\right]=\frac{2\left(\tilde{g}_{d}+\tilde{g}_{os}\right)^{2}\left(1+2\bar{n}_{-}\right)}{\omega-2E_{-}+i\kappa_{-}}.\label{eq:self2}
\end{align}
The cavity Green's function can then be directly found by applying the transformation in Eq.~(\ref{eq:trans}):
\begin{align} \label{GFaa}
G_{aa}^{R}\left[\omega\right]  = & V_{11}^{2}G^{R}_-\left[\omega\right]+ V_{12}^{2}G^{R}_+\left[\omega\right] \nonumber\\
& +V_{13}^{2}G^{R}_-\left[-\omega\right]^{*}+V_{14}^{2}G^{R}_+\left[-\omega \right]^{*} ,
\end{align}
from which we easily compute the cavity density of states as $\rho_a[\omega] = - {\rm Im}G_{aa}^{R}\left[\omega\right]/\kappa  $. Again, it is worth stressing that these formulas are mainly given here for reference, as they essentially coincide with those of Ref.~\cite{nonlinear1}, except for the replacement $\tilde{g}_{os} \to (\tilde g_{d}+\tilde g_{os})$ in Eqs.~(\ref{eq:self1}) and (\ref{eq:self2}). Still, the parametric dependence of these expressions turns out to be quite different from the case $\eta=0$. As we have already discussed, the presence of a large Duffing nonlinearity has significant qualitative and quantitative consequences on the behavior of the system, by influencing the photonic/phononic natures of the polaritons, the position of the resonant point, and the strength of the effective interaction between polaritons.

In general, the upper polariton is more sensitive to non-linear effects, as can be seen by considering the typical scales of the two self energies at the polariton peaks:\begin{align}
&\Sigma_-^R[E_-]\sim \left(\tilde{g}_{d}+\tilde{g}_{os}\right)^{2} \frac{\bar n_- - \bar n_+}{\kappa_+ +\kappa_-} , \label{sigma_estimate1}\\
&\Sigma_+^R[E_+]\sim \left(\tilde{g}_{d}+\tilde{g}_{os}\right)^{2} \frac{1+2 \bar n_-}{\kappa_-} .\label{sigma_estimate2}
\end{align}
 As discussed at the end of Sec.~\ref{sec:SYSTEMS-IN-POLARITON}, the occupation numbers $\bar{n}_\pm $ are small at low temperatures, giving $|\bar{n}_- - \bar{n}_+|\simeq 1/8$.  Therefore, occupation factors suppress $\Sigma_-^R$ by about an order of magnitude in the low-temperature regime. Another important point concerns the linewidths $\kappa_\pm$. From Fig.~\ref{fig:input-1-1}(a), we see that the denominator in Eq.~(\ref{sigma_estimate1}) is $\kappa_+ +\kappa_- \sim \kappa$. On the other hand, $\kappa_-$ can become much smaller than $\kappa$. This occurs at $\Delta > 2\Omega_m$, when the effective damping of the lower polariton satisfies $\kappa_- \ll \kappa $ and the small denominator  of Eq.~(\ref{sigma_estimate2}) leads to a further significant enhancement of $\Sigma_+^R$ compared to $\Sigma_-^R$. Indeed, the  conditions $\overline \eta \gg 1$, $\Delta > 2\Omega_m$, and $\omega \simeq E_+$ are the most favorable to observe the effect of the Duffing nonlinearity on $\rho_a[\omega]$. 

%%%%%%%%%%%%%%%%%%%%%%%%%%%%%%%%%%%%%%%%%%%%%%%%%%%
\begin{figure}
\begin{centering}
\includegraphics[width=0.48\textwidth]{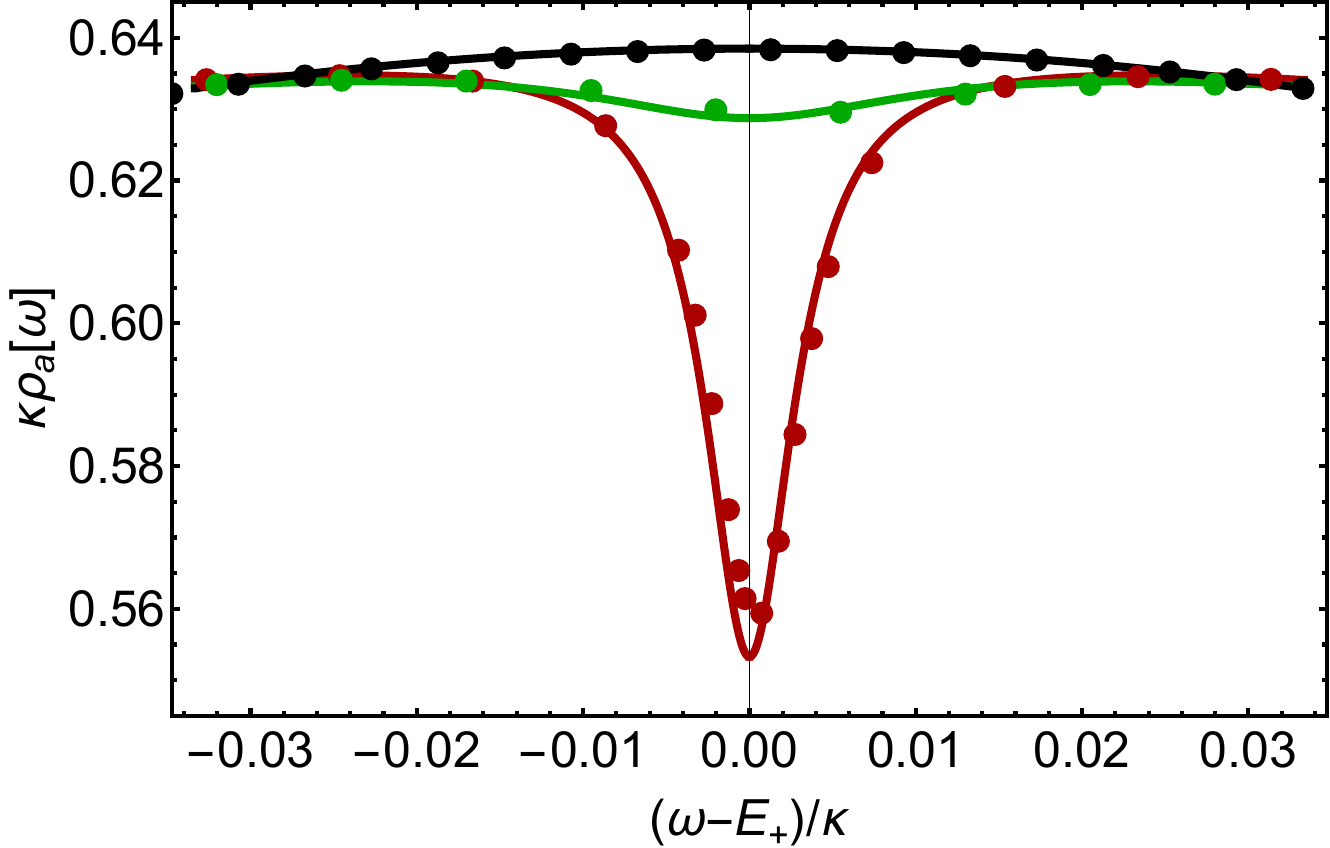}
\par\end{centering}
\caption{Suppression of cavity DOS at the upper polariton peak. Solid lines are obtained using Eq.~(\ref{GFaa}) and dots are from numerical simulation results based on a Lindblad master equation of the polaritons~\cite{nonlinear1,Lemode_PRA_2015,lijing2018,qiujing2022}. The three curves are obtained with $\eta/\Omega_{m}=8 \times 10^{-11}$ (black), $6\times10^{-9}$ (green), and $10^{-7}$ (red). Other parameters are $\kappa/\Omega_{m}=10^{-3}$, $g/\Omega_{m}=10^{-5}$, $\gamma/\Omega_{m}=10^{-7}$, $k_BT=0$, and  $\Delta/\Omega_{m}=3$, giving $\overline \eta =0.8, 60, 10^3$. \label{fig:desity}}
\end{figure}
%%%%%%%%%%%%%%%%%%%%%%%%%%%%%%%%%%%%%%%%%%%%%%

Specializing ourselves to this region of parameters, we show in Fig.~\ref{fig:desity} the effect of increasingly large values of the Duffing nonlinearity on $\rho_a[\omega]$, close to the frequency $\omega=E_+$ of the upper polariton. We see that the effect becomes clearly visible when entering the regime $\overline \eta \gg 1$ of Eq.~(\ref{large_eta}). Instead, the modification of the lineshape around $\omega  = E_-$ (not shown) is negligible with the parameters of Fig.~\ref{fig:desity}. We also observe that the effect of $\Sigma_+^R[\omega]$ is a narrow feature on top of a much broader peak~\cite{nonlinear1,teuful2013}. This is again in contrast to the behavior at the lower polariton, for which the linewidth of the self-energy is determined by  $\kappa_++\kappa_-$, see Eq.~(\ref{eq:self1}). Since $\kappa_+ +\kappa_- > \kappa_-$ (i.e., larger than the linewidth of the polariton peak itself), the self-energy $\Sigma_-^R[\omega]$ always gives a broad modification of the lineshape, without introducing qualitatively striking features. A similar remark holds for the upper polariton in the range of detunings $\Omega_m/2 <\Delta < 2\Omega_m$. In this case, as seen in Fig.~\ref{fig:input-1-1}(a), the linewidth $\kappa_-$ of $\Sigma_+^R[\omega]$ is much larger than  $\kappa_+$, thus the self-energy can be simply approximated as a constant.

Before analyzing in detail the parameter dependence of the dip shown in Fig.~\ref{fig:desity}, we briefly discuss the validity of the perturbative treatment. The zero-order self-energy of the higher polariton is simply $\kappa_+$, so a conservative condition to apply perturbation theory reads $\Sigma_+ \ll \kappa_+$ which, at low temperature, can be written $C_{eff}\sim \Sigma_+/\kappa_+ \ll 1$ [for the precise definition of $C_{eff}$, see Eq.~(\ref{eq:ceff})]. Finally, as we are interested in the regime of large Duffing nonlinearity, $C_{eff}$ can be estimated using Eq.~(\ref{Ceff_estimate_Duffing}). This gives the following upper bound for $\eta$:
\begin{equation}\label{perturbative_condition}
\eta \ll \frac{\kappa^2}{\Omega_m}.
\end{equation} 
Note that the above condition is consistent with Fig.~\ref{fig:desity}. The red analytical curve corresponds to the largest value of $\eta = 0.1 \kappa^2/\Omega_m$ and starts to show small deviations from the non-perturbative numerical results. Instead, excellent agreement is found for the curves with smaller $\eta$.

\subsection{Nonlinear signature for $\Delta>2\Omega_{m}$}

Following the previous arguments, we restrict ourselves to the upper polariton peak and first analyze quantitatively the most promising range of detuning, $\Delta>2\Omega_{m}$. The suppression of cavity DOS can be conveniently characterized by the effective cooperativity~\cite{nonlinear1}:
\begin{equation}\label{eq:ceff}
C_{eff}=\frac{4\left(\tilde{g}_{d}+\tilde{g}_{os}\right)^{2}\left(1+2\bar{n}_{-}\right)}{\kappa_{+}\kappa_{-}},
\end{equation}
giving
\begin{equation}
\rho_{a}\left[E_{+}\right] \simeq \frac{2}{\pi\kappa_{+}}\left(1+C_{eff}\right)^{-1}.
\end{equation}
In other words, the cavity DOS is reduced by a factor $1+C_{eff}$ where, typically, $C_{eff} \ll 1$. The main objective of this subsection is to obtain analytic approximations of $C_{eff}$, from which we will be able to characterize the optimal operation point.

Useful expressions of the various quantities entering Eq.~(\ref{eq:ceff}) can be obtained by performing an expansion in powers of the (small) bare optomechanical coupling $g$. To leading order, we obtain:
\begin{align}
&\frac{\tilde{g}_{os}}{g} \simeq \frac{\sqrt{\frac{\Omega_m}{\Delta} \left(1 +\frac{4\Omega_m^2}{\Delta^2} -\frac{32 \Omega_m^4}{\Delta^4}\right)}}{\left( 54 \overline\eta\right)^{1/4}\left( 1- \frac{4\Omega_m^2}{\Delta^2}\right)^{1/4}},  \\
&\frac{\tilde{g}_d}{g} \simeq  \left( \frac{ 2\overline\eta}{243} \right)^{\frac14}\left(1-\frac{4\Omega_m^2}{\Delta^2}\right)^\frac34\sqrt{\frac{\Omega_m}{\Delta}\left(1+\frac{8\Omega_m^2}{\Delta^2}\right)}.
\end{align}
From above, and assuming $\Delta \sim \Omega_m$, the effective couplings are estimated as follows:
\begin{align} \label{eff_g_scales} 
\tilde{g}_{os} \sim \frac{g} {\overline\eta^{1/4}},  
\quad {\rm and } \quad
\tilde{g}_d \sim  g \overline\eta^{1/4} .
\end{align}
As expected, $\tilde{g}_d$ becomes the dominant nonlinear coupling at large $\eta$. However, the above dependence might not be completely intuitive. In particular, while $\tilde{g}_d$ is directly related to the strength of the Duffing nonlinearity, its dependence is not simply proportional to $\eta$. This is because the effective couplings are also strongly influenced by the degree of hybridization of the polaritons and the value of $\beta$. 

To understand the dependence of the effective couplings on $\eta$, we should take into account that  at large $\eta$ the resonant polaritons are weakly hybridized. An explicit calculation gives:   
\begin{align}
\frac{G_{s}}{\Omega_{m}} & \simeq \frac{\sqrt{ \Delta^2+8 \Omega_m^2}}{ 2\Omega_m  (216 \overline\eta)^{1/4} }\left(1-\frac{4\Omega_m^2}{\Delta^2}\right)^\frac14 \sim \frac{1}{\overline\eta^{1/4}},
\end{align}
showing that the dressed optomechanical coupling $G_s$ is typically much smaller than the difference between the bare energies $\Delta,\widetilde\Omega_m$ of the non-interacting modes. The weak hybridization has a negative impact on the effective couplings $\tilde g_{os}$ and $\tilde g_{d}$. For example, considering the definition of  $\tilde g_{os}$ in Eq.~(\ref{eq:optonon}) allows us to trace the suppression of the bare optomechancal coupling $g$ to the small matrix element $V_{11} \sim \overline \eta^{-1/4}$ (while $|V_{21}|\sim|V_{12}| \sim 1$), obtaining a result in agreement with Eq.~(\ref{eff_g_scales}). For the nonlinear coupling $\tilde{g} _d$ we have instead $\tilde g _d \sim \eta \beta( V_{22}+V_{24})$, where $V_{22},V_{24}$ scale as $\overline \eta^{-1/4}$. Furthermore, Eq.~(\ref{eq:beta_res2}) gives $\beta \sim \sqrt{\Omega_m/\eta} $. We see that a stiffer nonlinearity contributes negatively to $\tilde g _d $ by reducing the equilibrium position of the mechanical element. These two effects (weak hybridization and smaller $\beta$)  partially compensate the linear growth induced by the coupling coefficient $\eta$, to give a final estimate of $\tilde g _d $ in agreement with Eq.~(\ref{eff_g_scales}). 

%%%%%%%%%%%%%%%%%%%%%%%%%%%%%%%%%%%%%%%%%%%%%%%
\begin{figure}
\begin{centering}
\includegraphics[width=0.45\textwidth]{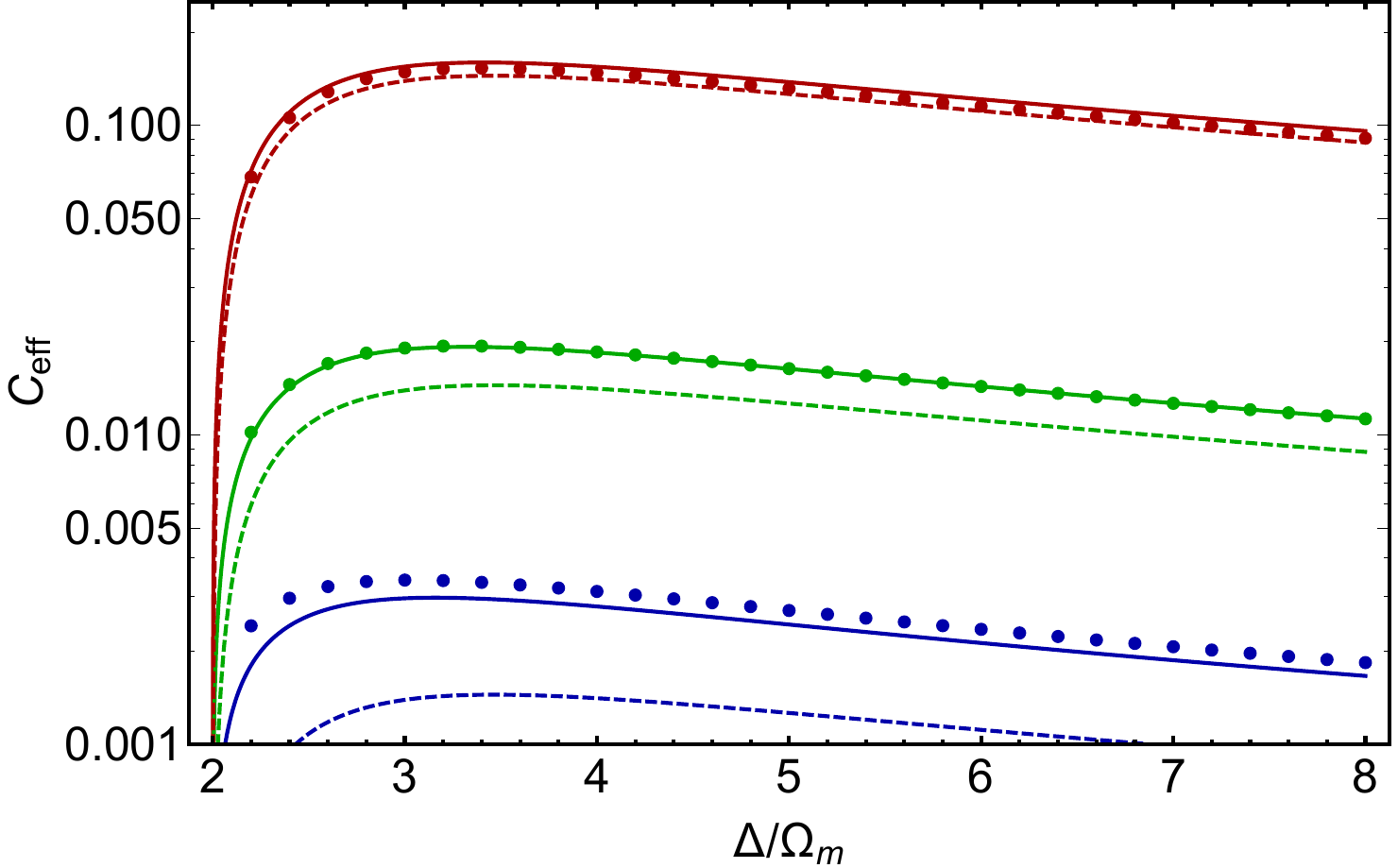}
\par\end{centering}
\caption{Dependence of $C_{eff}$ on detuning in the region $\Delta > 2\Omega_m$. Numerical data (dots) are obtained from Eq.~(\ref{eq:ceff}) using $\eta/\Omega_m =10^{-9}$ (blue), $10^{-8}$ (green), and $10^{-7}$ (red). The dashed and solid curves are plots of Eqs.~(\ref{Ceff_leading_order}) and (\ref{Ceff_next_order}), respectively. Other parameters are $\kappa/\Omega_m=10^{-3}$, $g/\Omega_m=10^{-5}$, and
$\gamma/\Omega_m=10^{-7}$ .\label{fig:ceff1}}
\end{figure}
%%%%%%%%%%%%%%%%%%%%%%%%%%%%%%%%%%%%%%%%%%%%%%%

We now consider the other  quantities entering Eq.~(\ref{eq:ceff}). For the lower polariton damping rate we get:
\begin{align} \label{kappa_m_estimate}
\kappa_- \simeq  \frac{2 \kappa }{27} \sqrt{\frac{2}{3\overline \eta}\left(1-\frac{4\Omega^2}{\Delta^2}\right)} \left(1+\frac{8\Omega_m^2}{\Delta^2}\right) \sim\frac{\kappa}{\overline \eta^{1/2}}.
\end{align}
This expression was derived setting $\gamma=0$, which is numerically accurate if $\overline\eta$ is not exceedingly large. Instead, the upper polariton has $\kappa_+ \simeq \kappa$. Finally, to leading order the effective occupation is $\bar n_- \simeq 1/8 $, giving:
\begin{equation} \label{Ceff_leading_order}
C_{eff} \simeq \overline\eta \frac{15 g^2 }{2  \kappa^2} \frac{\Omega_m}{\Delta} \left(1-\frac{4\Omega_m^2}{\Delta^2}\right)  \sim \overline\eta \frac{g^2 }{\kappa^2} .
\end{equation}
We note that the value $C_{eff} \sim g^2/\kappa^2$, which is typically attained in the absence of Duffing nonlinearity, gets enhanced by the $\overline \eta$ prefactor.

Plots of $C_{eff}$ as a function of $\Delta$ are shown in Fig.~\ref{fig:ceff1}, where we compare Eq.~(\ref{Ceff_leading_order}) to the numerical evaluation of Eq.~(\ref{eq:ceff}). Unfortunately, significant deviations appear when considering smaller values of $\eta$, which makes it necessary to include the subleading term. By a tedious but relatively straightforward calculation we obtain:
\begin{align} \label{Ceff_next_order}
 C_{eff}  =   & \frac{\overline\eta  g^2 }{\kappa^2}  \frac{\Omega_m}{\Delta}
\left[\frac{15 }{2  }\left(  1-\frac{4\Omega_m^2}{\Delta^2}\right) +\frac{\overline\eta^{-1/2}}{36\sqrt{6} }\sqrt{ 1-\frac{4\Omega_m^2}{\Delta^2}} \right. \nonumber \\
&\left. \times\left(1811-796\frac{\Omega_m^2}{\Delta^2}+3008\frac{\Omega_m^4}{\Delta^4}\right)+ \ldots\right],
\end{align}
where the second term in the square parenthesis, suppressed by a factor $\overline \eta^{-1/2}$, improves considerably the agreement with the numerical values. If desired, the expansion can be pursued further. The next-order term has a rather cumbersome expression, which we report explicitly in Appendix~\ref{app:expansion}.

\subsection{Nonlinear signature for $\Omega_m/2<\Delta<2\Omega_{m}$}

A similar analysis can be performed in the interval  $\Omega_m/2<\Delta<2\Omega_{m}$, where we obtain:
\begin{align}
&\frac{\tilde{g}_{os}}{ g} \simeq \frac{  1}{18 \sqrt{3 \overline  \eta} } \sqrt{\frac{\Omega_m}{ \Delta}\left( 4-\frac{\Omega_m^2}{\Delta^2}\right)} \left( 2+\frac{\Omega_m^2}{\Delta^2} \right) ,  \\
&\frac{\tilde{g}_d}{g} \simeq  \frac{1}{27} \sqrt{\frac{2\Omega_m}{\Delta}} \left(8+\frac{2\Omega_m^2}{\Delta^2}-\frac{\Omega_m^4}{\Delta^4} \right).
\end{align}
Interestingly, although we still have $\tilde{g}_d \gg \tilde{g}_{os}$, the coupling $\tilde{g}_d$ does not increase with $\eta$ but remains of the same order of the bare optomechanical coupling $g$. The reason lies in a much weaker hybridization between cavity and mechanical modes. In this regime we have $V_{22} \simeq 1$, therefore Eq.~(\ref{eq:duffing}) gives $\tilde{g}_d \sim \eta \beta (V_{21}+V_{23})^2$. Similarly to the previous case, $\beta \sim \sqrt{\Omega_m/\eta}$ and $V_{21},V_{23} \sim \overline \eta^{-1/4}$. However, the matrix elements are squared in this case, and this results in the cancellation of the overall dependence on $\eta$. The absence of an enhancement factor for the effective nonlinear couplings impacts negatively the value of $C_{eff}$ in this regime.

To leading order, the damping for the upper polariton is given by:
\begin{align}
\kappa_+ \simeq \frac{2  \kappa }{27}
\sqrt{\frac{8}{3\overline\eta}\left(1 - \frac{\Omega_m^2}{4\Delta^2}\right)}
\left(2 + \frac{\Omega_m^2}{\Delta^2}\right),
\end{align}
which has a structure similar to Eq.~(\ref{kappa_m_estimate}). Instead the lower polariton has a photonic character, with $\kappa_- \simeq \kappa$. Finally, the effective occupation of the lower polariton is generally small:
\begin{align}
\bar n_- \simeq \frac{8}{243 \overline \eta} \left( 1 + \frac{3\Omega_m^2}{4\Delta^2}-\frac{\Omega_m^6}{16\Delta^6} \right).
\end{align}
From the above expressions, we obtain the leading-order contribution to $C_{eff}$:
\begin{equation} \label{Ceff_leading_order_2}
C_{eff} \simeq \overline\eta^{1/2} \frac{2\sqrt{6} g^2 }{27  \kappa^2} \frac{\Omega_m}{\Delta}\left(2+\frac{\Omega_m^2}{\Delta^2}\right)\left(4-\frac{\Omega_m^2}{\Delta^2}\right)^{3/2},
\end{equation}
which is only proportional to $\overline\eta^{1/2}$. As noted already, the difference from Eq.~(\ref{Ceff_leading_order}), where $C_{eff}\propto \overline\eta$, is  due to the weak dependence of $\tilde g_d$ on $\overline\eta$. Therefore, in the range $\Omega_m <\Delta < 2\Omega_m$ the enhancement of $C_{eff}$ is entirely due to the suppression of $\kappa_+$ with $\overline\eta$.

\begin{figure}
\begin{centering}
\includegraphics[width=0.45\textwidth]{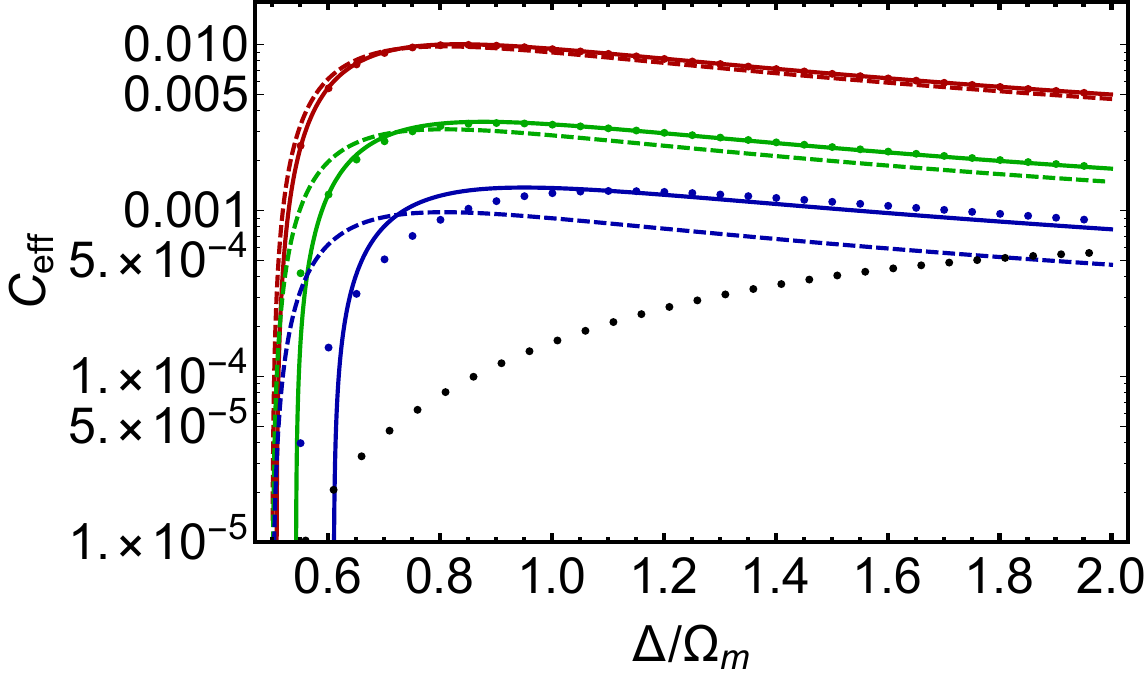}
\par\end{centering}
\caption{Dependence of $C_{eff}$ on detuning in the region $\Omega_m/2 < \Delta < 2\Omega_m$. Numerical data (dots) are obtained from Eq.~(\ref{eq:ceff}) using $\eta/\Omega_m =0 $ (black), $\eta/\Omega_m =10^{-9}$ (blue), $10^{-8}$ (green), and $10^{-7}$ (red). The dashed and solid curves are plots of Eqs.~(\ref{Ceff_leading_order_2}) and (\ref{Ceff_next_order_2}), respectively. Other parameters are $\kappa/\Omega_m=10^{-3}$, $g/\Omega_m=10^{-5}$, and
$\gamma/\Omega_m=10^{-7}$.\label{fig:ceff2}}
\end{figure}

A comparison of numerical and analytical results is shown in Fig.~\ref{fig:ceff2}, where we also plot the value of $C_{eff}$ with $\eta = 0$. The presence of a strong Duffing nonlinearity allows to enhance the value of $C_{eff}$. However, a comparison to Fig.~\ref{fig:ceff1} shows that the case of larger detunings $\Delta > 2\Omega_m$ is much more favorable. Again, with smaller values of $\eta$ relatively large deviations appear between the numerical results and Eq.~(\ref{Ceff_leading_order_2}). A more accurate expression, including the subleading term, reads:
\begin{align} \label{Ceff_next_order_2}
C_{eff} \simeq  &  \overline\eta^{1/2} \frac{2g^2}{27\kappa^2} \frac{\Omega_m}{\Delta}\left[ \sqrt{6}\left(2+\frac{\Omega_m^2}{\Delta^2}\right)\left(4-\frac{\Omega_m^2}{\Delta^2}\right)^{3/2}\right. \nonumber \\
& \left.  +\frac{2056+770\frac{\Omega_m^2}{\Delta^2}-1209\frac{\Omega^4}{\Delta^4}+38\frac{\Omega_m^6}{\Delta^6}+46\frac{\Omega_m^8}{\Delta^8}}{27 \overline\eta^{1/2}} \right].
\end{align}

%%%%%%%%%%%%%%%%%%%%%%%%%%%%%%%%%%%%%%%%%%%%%%%
%%%%%%%%%%%%%%%%%%%%%%%%%%%%%%%%%%%%%%%%%%%%%%%

%%%%%%%%%%%%%%%%%%%%%%%%%%%%%%%%%%%%%%%%%%%%%%%
\subsection{Optimal point}
%%%%%%%%%%%%%%%%%%%%%%%%%%%%%%%%%%%%%%%%%%%%%%%

We have established that, in the presence of a strong Duffing nonlinearity, the regime $\Delta > 2\Omega_m$ of Fig.~\ref{fig:ceff1} is generally more favorable. Interestingly, Fig.~\ref{fig:ceff1} also shows that the dependence on $\Delta$ is non-monotonic, thus there is an optimal value of the detuning. From Eq.~(\ref{Ceff_leading_order})  we immediately obtain that, to leading order, the optimal detuning is $\Delta_{opt} \simeq 2\sqrt{3}\Omega_m$, giving $C_{eff}^{opt} \simeq  5\sqrt{3}\overline \eta g^2/6\kappa^2$. Reintroducing the physical coupling $\eta$, we can estimate
\begin{equation}
    C_{eff}^{opt} \simeq  \frac{5\sqrt{3}\eta \Omega_m}{6\kappa^2},
\end{equation}
which is independent on the optomechanical coupling $g$. More accurate analytical expressions, including the higher-order terms, read as follows:
\begin{align}\label{delta_opt_better}
\Delta_{opt} \simeq 2\sqrt{3}\Omega_m \left(1-\frac{2501}{4860 \sqrt{\overline\eta}} + \frac{4321876}{7381125 \overline\eta }  \right),
\end{align}
and
\begin{align}\label{Ceff_opt_better}
C_{eff}^{opt} \simeq \overline\eta  \frac{5\sqrt{3} g^2}{6\kappa^2} \left(1+\frac{ 1589}{486  \sqrt{\overline\eta}}  + \frac{ 644327}{209952  \overline\eta }  \right).
\end{align}
The above equations cannot be obtained from the previous Eq.~(\ref{Ceff_next_order}), which only includes the first subleading correction. Instead, to derive them, we have also taken into account the $n=3$ term of  Eq.~(\ref{Ceff_general_expansion}). Figure~\ref{fig:copt} shows that Eq.~(\ref{Ceff_opt_better}) describes well the growth of $C_{eff}^{opt}$ when $\overline \eta >1$.  From Fig.~\ref{fig:ceff2} we see that a non-monotonic dependence is also found when $\Omega_m/2< \Delta < 2\Omega_m$. In that interval of detuning, the maximum is attained at $\Delta \simeq  0.81 \Omega_m$, giving
$C_{eff}^{opt}  \simeq 3.1 \overline\eta^{1/2} g^2/\kappa^2 $. As noted already,  the scaling with $\overline 
\eta$ is weaker in this case.

%%%%%%%%%%%%%%%%%%%%%%%%%%%%%%%%%%%%%%%%%%%%%%%%%%%%%%%%%%%%
\subsection{Effects of temperature}
%%%%%%%%%%%%%%%%%%%%%%%%%%%%%%%%%%%%%%%%%%%%%%%%%%%%%%%%%%%%

As seen in  Eq.~(\ref{eq:ceff}), a finite temperature further enhances $C_{eff}$ through the increase in population $\bar n_-$ of the lower polariton.  At sufficiently large values of $k_B T/\Omega_m$, the value of $\bar{n}_-$ is dominated by thermal fluctuations and, as in the case without Duffing nonlinearity, can be described by a classical treatment of the equations of motion~\cite{nonlinear1,teuful2013,Lemode_PRA_2015}. Instead, at low temperature, the occupation number is determined by vacuum fluctuations or, more accurately, by quantum activation induced by the optical drive (see the discussion at the end of Sec.~\ref{sec:SYSTEMS-IN-POLARITON}). In that limit, the OMIT signal cannot be described classically. Below, we discuss in detail when temperature effects start to play an important role, and their influence on the visibility of the OMIT dip.

%%%%%%%%%%%%%%%%%%%%%%%%%%%%%%%%%%%%%%%%%%%%
\begin{figure}
\begin{centering}
\includegraphics[width=0.49\textwidth]{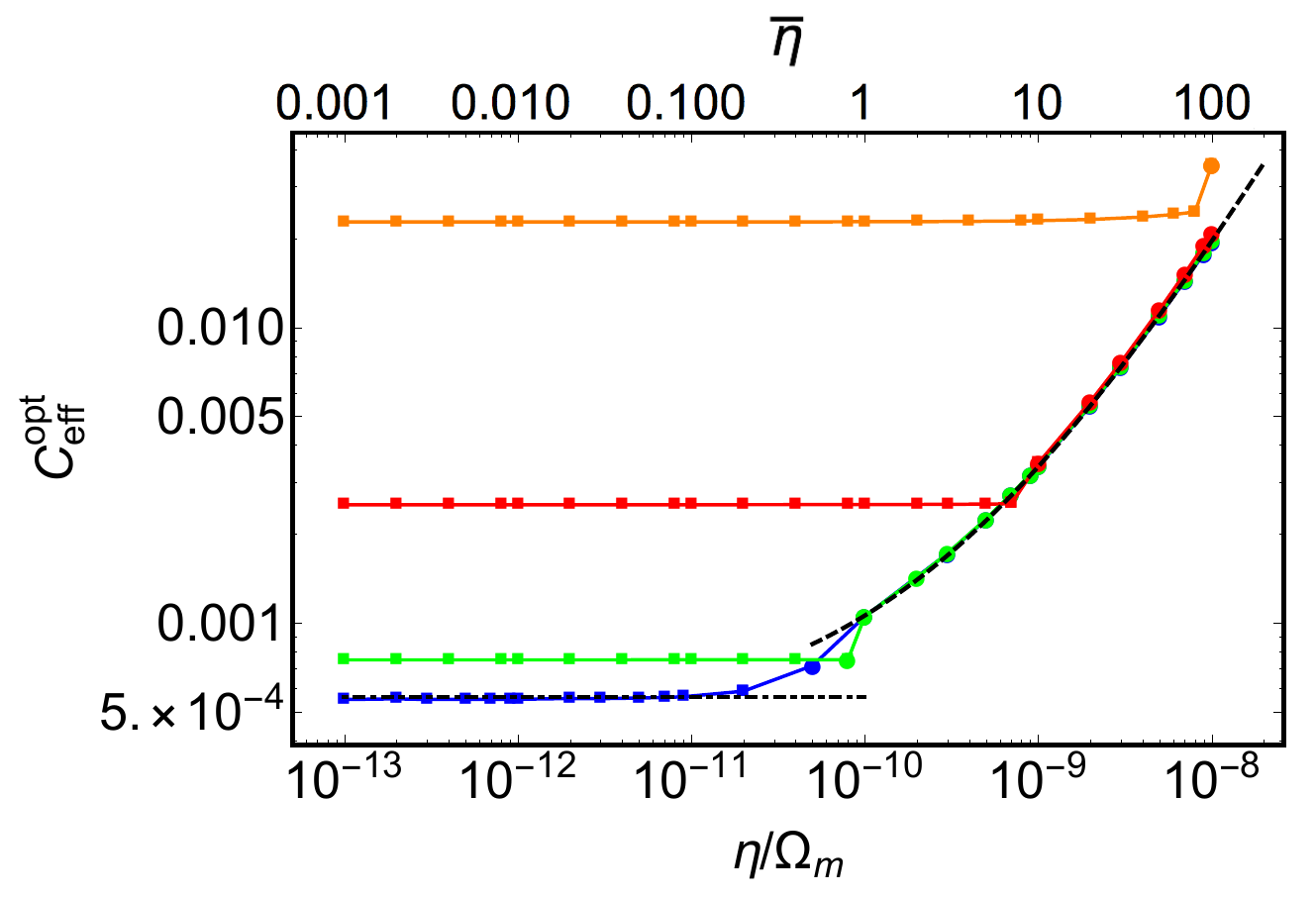}
\par\end{centering}
\caption{Dependence of $C^{opt}_{eff}$ on the strength $\eta$ of the Duffing nonlinearity. The four curves are obtained at increasing temperature, using $\bar n_{th}(\Omega_m)=0,2,10,100$ (from bottom to top). The square (round) markers indicate if the optimal point is obtained in the interval $\Omega_m/2 <\Delta <2\Omega_m$ ($\Delta> 2\Omega_m$). The asymptotic dependence of the zero-temperature case is given by $C_{eff}^{opt}= 45g^2/8\kappa^2$ when  $\eta\to 0$ (dot-dashed line) and by Eq.~(\ref{Ceff_opt_better}) when  $\overline\eta \gg 1$ (dashed curve). Other parameters are: $\kappa/\Omega_m=10^{-3}$, $g/\Omega_m=10^{-5}$, and
$\gamma/\Omega_m=10^{-7}$ .\label{fig:copt}}
\end{figure}
%%%%%%%%%%%%%%%%%%%%%%%%%%%%%%%%%%%%%%%%%%%%

We first show in Fig.~\ref{fig:copt} how the dependence of $C^{opt}_{eff}$ on $\eta$ is modified by a finite temperature. In the $\eta \to 0$ limit, the maximum of $C_{eff}$ is approximately given by $C^{opt}_{eff} \sim \bar n_{th}(\Omega_m)g^2/\kappa^2$~\cite{nonlinear1,teuful2013}. By increasing $\eta$, we find that this value of $C_{eff}$ remains approximately constant until the condition $\overline\eta \gtrsim \bar n_{th}(\Omega_m)$ is met. Then, $C^{opt}_{eff}$ grows with $\overline\eta$ and is weakly affected by temperature. In summary, we can estimate  $C^{opt}_{eff}$ as follows:
\begin{equation}\label{Copt_max_finiteT}
C^{opt}_{eff} \sim \max\left [\bar n_{th}(\Omega_m),\overline\eta \right] \frac{g^2}{\kappa^2},
\end{equation}
where the enhancement factor is determined by the dominant mechanism, either temperature or the large Duffing nonlinearity.

\begin{figure}
\begin{centering}
\includegraphics[width=0.22\textwidth]{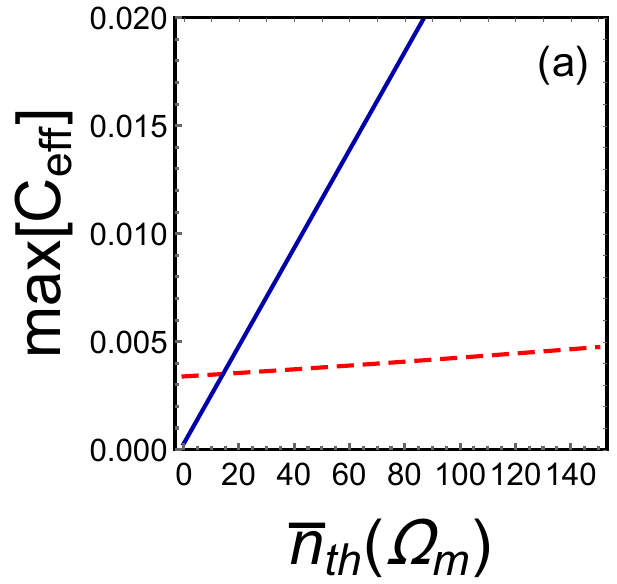}
\includegraphics[width=0.23\textwidth]{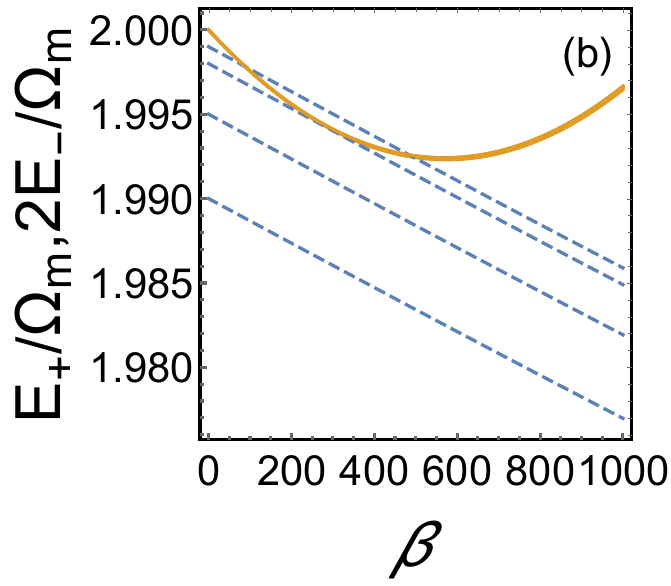}
\includegraphics[width=0.47\textwidth]{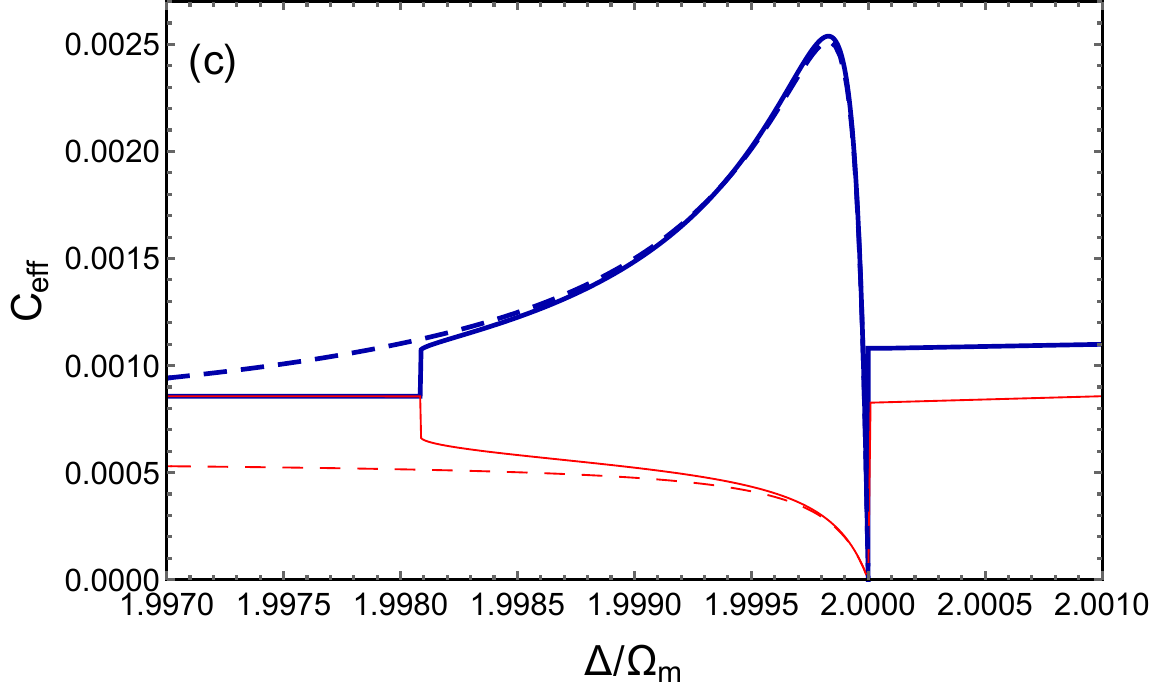}
\par\end{centering}
\caption{ Detailed analysis of the temperature dependence shown in Fig.~\ref{fig:copt}, for the specific value $\eta=10^{-9}\Omega_m$. Panel (a) shows the temperature dependence of the maximum $C_{eff}$ in the intervals $1.998 \Omega_m \leq \Delta < 2\Omega_m$ (solid curve) and $\Delta \geq 2\Omega_m$ (dashed curve). Panel (b) illustrates the appearance of additional solutions to Eq.~(\ref{resonance_condition}) in the interval $1.998 \Omega_m \leq \Delta < 2\Omega_m$. The dashed curves show the dependence of $2E_-$ on $\beta$ for different values of the detuning $\Delta/\Omega_m=1.99,1.995,1.998,1.999$ (from bottom to top). The solid curves, which overlap for the same detuning values, refer to the upper polariton $E_+$. Panel (c) illustrates the dependence of $C_{eff}$ on detuning around the singular point $\Delta = 2\Omega_m$, for $\bar{n}_{th}(\Omega_m)=10$ (thick blue curves) and $\bar{n}_{th}(\Omega_m)=0$ (thin red curves). The solid (dashed) curves refer to $\eta=10^{-9}\Omega_m$ ($\eta=0$). Other parameters are as in Fig.~\ref{fig:copt}: $\kappa/\Omega_m=10^{-3}$, $g/\Omega_m=10^{-5}$, and
$\gamma/\Omega_m=10^{-7}$ .\label{fig:Tpeak}}
\end{figure}

It is interesting to discuss in more detail how the behavior of Eq.~(\ref{Copt_max_finiteT}) arises, as it is not simply due to a thermal enhancement of $\bar{n}_-$ around the zero-temperature optimal point. If this were the case, the effect of temperature would be much weaker. In fact, using Eq.~(\ref{eq:nn}), the occupation number can be estimated as follows:
\begin{equation}
\bar{n}_- \sim \max\left[\frac18,\sqrt{\bar\eta}\frac{\gamma}{\kappa} \frac{k_B T}{\Omega_m} \right].
\end{equation}
In the above expression, the value 1/8 is the $T\to 0$ limit and the large-temperature dependence is obtained from the first term in the outer parentheses of Eq.~(\ref{eq:nn}), using $V_{21}\sim 1$ and $\kappa_-\sim \kappa\bar\eta^{-1/2}$. We see that the growth of the polariton number with temperature is suppressed by a prefactor of order $\bar{\eta}^{1/2}\gamma/\kappa$, which is usually small. Indeed, as shown by the dashed curve in Fig.~\ref{fig:Tpeak}(a), if one restricts the maximization of $C_{eff}$ to the interval $\Delta>2\Omega_m$, the dependence of $C_{eff}^{opt}$ on temperature would be rather weak and, in particular, much weaker than the temperature dependence for detunings slightly below $\Delta= 2\Omega_m$ (solid curve).

%%%%%%%%%%%%%%%%%%%%%%%%%%%%%%%%%%%%%%%%%%%%
%%%%%%%%%%%%%%%%%%%%%%%%%%%%%%%%%%%%%%%%%%%% 

As it turns out, Eq.~(\ref{Copt_max_finiteT}) is determined by the appearance of additional solutions of the resonant condition. Previously we have explained that, as shown in panels (a) and (b) of Fig.~\ref{fig:envsbeta}, there is usually a single solution of Eq.~(\ref{resonance_condition}) when $\Delta < 2 \Omega_m$. In reality, however, multiple solutions can be found in a tiny region below $\Delta=2\Omega_m$. For specific system parameters, the appearance of these additional solutions is shown explicitly in Fig.~\ref{fig:Tpeak}(b). At sufficiently low temperature, these solutions can be discarded as they do not affect the optimal $C_{eff}$. They appear only in a very small range of detunings ($1.998<\Delta/\Omega_m <2$ for the case of Fig.~\ref{fig:Tpeak}), where the value of $C_{eff}$ is dominated by the optomechanical nonlinearity. In Fig.~\ref{fig:Tpeak}(c), we compare the cases with (solid curves) and without (dashed curves) Duffing nonlinearity, finding excellent agreement in the interval $1.998<\Delta/\Omega_m <2$ both at finite temperature (thick curves) and at $T=0$ (thin curves). Therefore the effective cooperativity is given by $C_{eff}\sim g^2/\kappa^2$ at low temperature, which is much smaller than Eq.~(\ref{Ceff_estimate_Duffing}).

The situation reverses at sufficiently large temperature, as the value of $C_{eff}$ around $\Delta=2\Omega_m$ gets strongly influenced and becomes dominant. This can be seen in Fig.~\ref{fig:Tpeak}(b), where the maximum value of the effective cooperativity in the interval $1.998<\Delta/\Omega_m <2$, approximately given by $C_{eff}\sim \bar{n}_{th} (\Omega_m)g^2/\kappa^2$, grows much faster than in the region $\Delta>2\Omega_m$. It is this maximum at $\Delta\simeq 2\Omega_m$ which determines the high-temperature limit of Eq.~(\ref{Copt_max_finiteT}).

\subsection{Experimental parameters}\label{exp_parameters}

\begin{figure}
\begin{centering}
\includegraphics[width=0.43\textwidth]{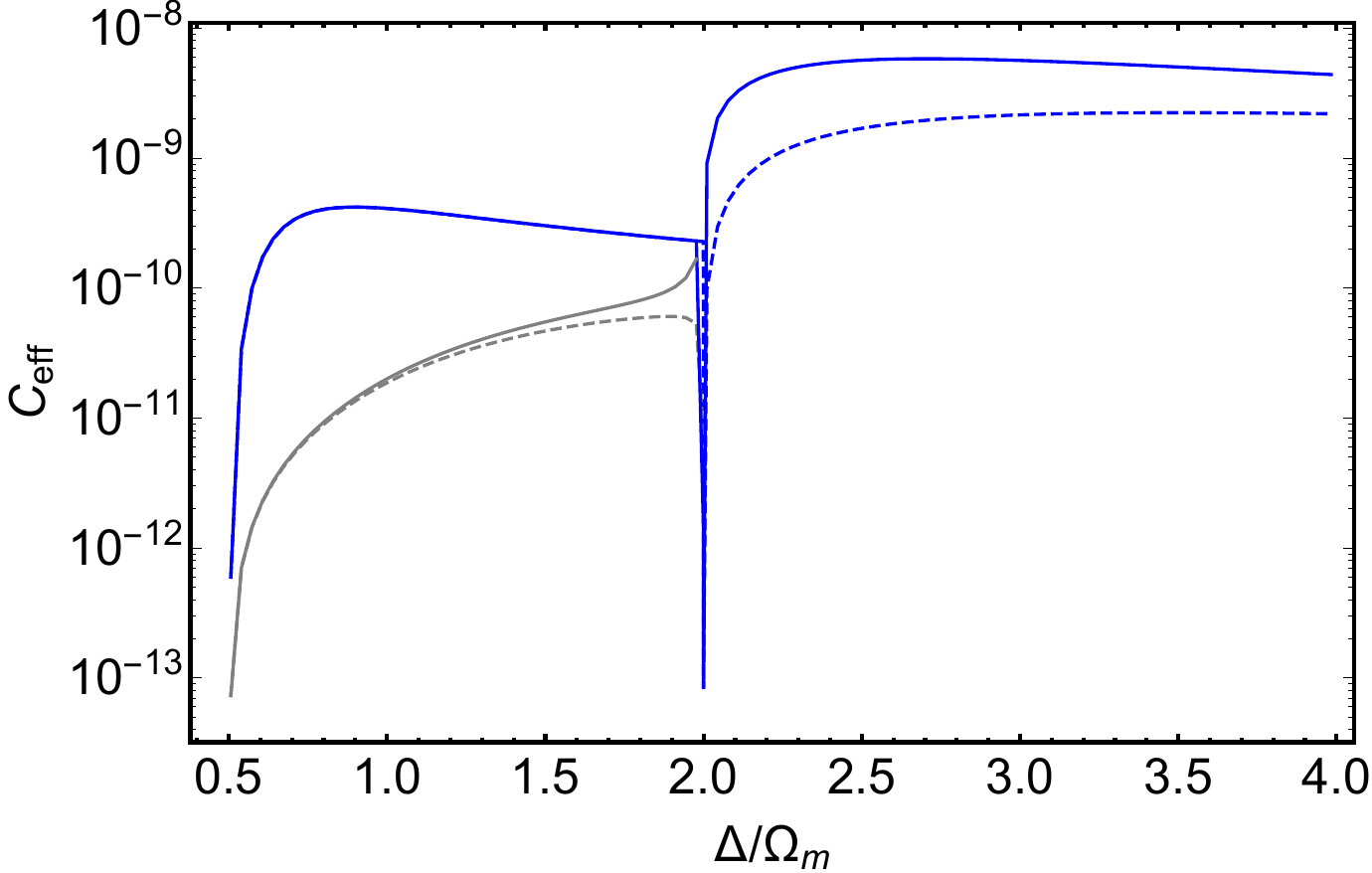}
\par\end{centering}
\caption{Dependence of $C_{eff}$ on detuning, with parameters taken from Ref.~\cite{duffingex4}: $\Omega_m =2\pi \times 36.2~{\rm MHz}$, $\kappa=2\pi \times 242~{\rm kHz}$, $g=2\pi\times 0.83~{\rm Hz}$, and $\gamma= 2\pi \times 700~{\rm Hz}$. The two upper blue curves use $\eta = 2\pi \times 2.4~{ \mu\rm Hz}$~\cite{duffingex4}, while the lower two lower gray curves assume $\eta=0$. The temperature is either $T=0$ (dashed) or 14~mK (solid). \label{fig:ceff_exp}}
\end{figure}

In experiments, relatively large Duffing nonlinearities have been realized with mechanical resonators formed by single- or multi-layer graphene~\cite{duffingex3,Singh_NNano_2014,duffingex4}. The Duffing parameter $\alpha \simeq 2.3\times 10^{15}~{\rm kg}~{\rm m}^{-2}{\rm s}^{-2}$ of the optomechanical setup  in Ref.~\cite{duffingex4} can be related to the coupling $\eta$ through the zero-point fluctuations $x_{zpf}=\sqrt{\hbar/2m\Omega_m}$, where $m$ is the effective mass of the mechanical mode. Using $m\simeq 0.28~{\rm pg} $ and $\Omega_m =2\pi \times 36.2~{\rm MHz}$~\cite{Singh_NNano_2014,duffingex4} we obtain $\eta = \alpha x_{zpf}^4/\hbar \simeq 2\pi \times 2.4~{\mu \rm Hz}$. This value, together with other parameters from Ref.~\cite{duffingex4}, is used in Fig.~\ref{fig:ceff_exp} to evaluate $C_{eff}^{opt}$, finding much larger values than the $\eta=0$ limit (gray curves). The estimate $C_{eff}^{opt} \sim \eta \Omega_m/\kappa^2 \simeq  10^{-9} $, as well as the enhancement factor $\overline\eta \simeq 126$, are in good agreement with Fig.~\ref{fig:ceff_exp}. This plot also shows the dependence of $C_{eff}$ on environment temperature. Notice that we have neglected nonlinear damping effects, which as shown in Ref.~\cite{duffingex4} may appear when mechanical amplitude becomes large. 

%%%%%%%%%%%%%%%%%%%%%%%%%%%%%%%%%%%%%%%%%%%%%%%
%%%%%%%%%%%%%%%%%%%%%%%%%%%%%%%%%%%%%%%%%%%%%%%

Despite the beneficial effect of the Duffing term, the nonlinear effects in Fig.~\ref{fig:ceff_exp} are still extremely small. This, however, does not appear to be a fundamental limitation. Single-layer graphene resonators with a much larger value 
of $\alpha = 1.4\times 10^{16}~{\rm kg}~{\rm m}^{-2}{\rm s}^{-2}$ have been realized~\cite{duffingex3} which, using the corresponding values $\Omega_m \simeq 2\pi\times 200$~MHz and $m \simeq 3.9\times 10^{-19}$~kg, give an energy scale $\eta \simeq 2\pi \times 0.24$~Hz. Embedding such a resonator in an optomechaincal setup with the same value of $\kappa$ would lead to an effective cooperativity of order $C_{eff}^{opt} \sim \eta \Omega_m/\kappa^2 \simeq  10^{-3}$. A similar estimate of $C_{eff}^{opt}$ is obtained for the carbon nanotube resonator of Ref.~\cite{duffingex3}, where $\alpha = 6\times 10^{12}~{\rm kg}~{\rm m}^{-2}{\rm s}^{-2}$,  $m \simeq 7.9\times 10^{-21}$~kg, and $\Omega_m \simeq 2\pi\times 250$~MHz, giving $\eta \simeq 2\pi \times 0.16 $~Hz. Again, by assuming $\kappa \simeq 2\pi\times 240$~kHz, we obtain  $C_{eff}^{opt} \sim  10^{-3}$. 

As in the case without Duffing nonlinearity, the above estimates can be further improved by reducing $\kappa$. It is also worth mentioning that proposals to introduce the Duffing nonlinearity by coupling the mechanical resonator to a qubit~\cite{nonge1,nori2015} suggest that much larger values of $\eta$ might be achievable, e.g., of order $\eta \sim 2\pi \times 0.2$~kHz~\cite{nori2015}, which would make the nonlinear effects predicted here easy to observe.

\section{\label{sec:Conclusion}Conclusion}
We have studied the signatures of a strong Duffing nonlinearity on OMIT, going beyond the classical treatment of the mechanical oscillator~\cite{2010_PRA_Agarwal,duffingex2,nonlinear2015} and accounting for the interplay of the optomechanical and Duffing nonlinear terms in the resonant scattering process between polaritons. In particular, we have identified the regime in which the presence of a strong Duffing coupling can greatly enhance the unique signatures of nonlinear effects induced by the optomechanical interaction alone~\cite{nonlinear1,florian2013,teuful2013}, possibly bringing them to experimental reach. 

Although the linear optomechanical interaction is necessary to establish such effects, the hybridization between optical and mechanical modes remains small at the optimal resonant point. Therefore, the effective interaction between polaritons might lead to practical schemes for the generation of nonclassical mechanical states, e.g., analogous to the reservoir engineering approach of Ref.~\cite{nongaussian2}. We also expect that, as an alternative to the Duffing nonlinearity, similar effects would follow from a non-linearity of the cavity alone~\cite{2021_PRR_Steele,2023_NatComm_Steele,Cai_2025}, which represents another interesting direction for future studies.

\section*{Acknowledgments}

We acknowledge enlightening discussions with G. C. La Rocca. S.C. acknowledges support from the Innovation Program for Quantum Science and Technology (Grant No.~2021ZD0301602) and the National Science Association Funds (Grant No.~U2230402). Y.D.W. acknowledges support from the Fundamental and Interdisciplinary Frontier Research Priority Program of Chinese Academy of Sciences (Grant No. XDB0920000), NSFC (Grant No.12275331), and the Penghuanwu Innovative Research Center (Grant No. 12047503).

\section*{Data Availability}
The data that support the findings of this study are available from the authors upon reasonable request.

\appendix

\section{Transformation matrix to polaritons\label{sec:Appendix_Bogolubov}}

The linear Hamiltonian $\hat{H}_l$ can be diagonalized by a Bogoliubov transformation as in Eq.~(\ref{eq:trans}). The matrix elements of $V$ are given by:
\begin{align}\label{V_matrix}
&V_{11}=h_{+}\left(\frac{\Delta_{c}}{E_{-}}\right)\sin u,\quad V_{12}=-h_{+}\left(\frac{\Delta_{c}}{E_{+}}\right)\cos u, \nonumber\\
&V_{13}=h_{-}\left(\frac{\Delta_{c}}{E_{-}}\right)\sin u,\quad V_{14}=-h_{-}\left(\frac{\Delta_{c}}{E_{+}}\right)\cos u, \nonumber\\
&V_{21}=h_{+}\left(\frac{\widetilde{\Omega}_{m}}{E_{-}}\right)\cos u,\quad V_{22}=h_{+}\left(\frac{\widetilde{\Omega}_{m}}{E_{+}}\right)\sin u, \nonumber\\
&V_{23}=h_{-}\left(\frac{\widetilde{\Omega}_{m}}{E_{-}}\right)\cos u,\quad V_{24}=h_{-}\left(\frac{\widetilde{\Omega}_{m}}{E_{+}}\right)\sin u,
\end{align}
where $h_{\pm}(x)=\left(\sqrt{x}\pm\sqrt{1/x}\right)/2$, the mixing angle $u$ between photon and phonon modes is:
\begin{equation}
u=\frac12 \arccos\frac{\Delta_{c}^{2}-\widetilde{\Omega}_{m}^{2}}{\sqrt{\left(\widetilde{\Omega}_{m}^{2}-\Delta_{c}^{2}\right)^{2}+16G_{s}^{2}\Delta_{c}\widetilde{\Omega}_{m}}},\label{eq:deangle}
\end{equation}
and the polariton eigenfrequencies are:
\begin{equation}\label{eq:enp}
E_{\pm}=\sqrt{\frac{\Delta_{c}^{2}+\widetilde{\Omega}_{m}^{2}}{2}\pm \sqrt{\frac{\big(\Delta_{c}^{2}-\widetilde{\Omega}_{m}^{2}\big)^2}{4}+4G_{s}^{2}\Delta_{c}\widetilde{\Omega}_{m}}}.
\end{equation}

\section{ Transformed Hamiltonian \label{sec:bath}}

In this Appendix we provide further details on how to derive the transformed Hamiltonian $\hat{H}_s$. In the laboratory frame, the Hamiltonian is $\hat{H} =\hat{H}_0 +\hat{H}_{dr} + \hat{H}_{diss} $, where $\hat{H}_0$  and $\hat{H}_{dr}$ are given by Eqs.~(\ref{H0}) and (\ref{Hdr}), respectively, and  the interaction of cavity and mechanics with the environment is described by:
\begin{align}\label{eq:diss}
\hat{H}_{diss}=&\sum_{i}\omega_{c,i}\hat{A}_{i}^{\dagger}\hat{A}_{i}+\sum_{i}\omega_{m,i}\hat{B}_{i}^{\dagger}\hat{B}_{i} \nonumber \\ &+\hat{H}_{\kappa,int}+\hat{H}_{\gamma,int},
\end{align}
where $\hat{A}_{i}$ and $\hat{B}_{i}$ are annihilation operators
for the cavity and mechanical bath modes, respectively, with frequencies $\omega_{c,i}$ and $\omega_{m,i}$. The interaction terms are:
\begin{align}
&\hat{H}_{\kappa,int}=i\sqrt{\frac{\kappa}{2\pi\rho_{c}}}\sum_{i}\left(\hat{A}_{i}^{\dagger}-\hat{A}_{i}\right)\left(\hat{a}+\hat{a}^{\dagger}\right), \\
&\hat{H}_{\gamma,int}=i\sqrt{\frac{\gamma}{2\pi\rho_{m}}}\sum_{i}\left(\hat{B}_{i}^{\dagger}-\hat{B}_{i}\right)\left(\hat{b}+\hat{b}^{\dagger}\right), \label{Hint_gamma}
\end{align}
where $\kappa$ and $\gamma$
are the damping rates of the cavity and the mechanics, while $\rho_{c}$ and $\rho_{m}$
are the density of states of the cavity and mechanical baths, respectively. We consider
here a Markovian environment, so the damping rates and density of states are all taken as independent of frequency. 

 As usual, we first write $\hat H$ in a rotating frame at the laser frequency, defined by the unitary transformation $\hat{U}=\exp\left[-i\omega_l t\left(\hat{a}^{\dag}\hat{a}+\sum_i \hat{A}_{i}^{\dagger}\hat{A}_{i}\right)\right]$. We also displace the bosonic operators, to eliminate the driving terms from the equations of motion. This leads to Eqs.~(\ref{eq:steady1}) and (\ref{eq:steady2}) for the classical amplitudes and brings the Hamiltonian to the following form:
\begin{align}\label{Hd}
\hat{H}_{d}=&\Delta_c \hat{a}^\dag\hat{a} +\Omega_m \hat{b}^\dag\hat{b} +6 \eta \beta^2 \left(\hat{b}^{\dag}+\hat{b} \right)^2\nonumber \\
&-G\left(\hat{a}^{\dagger}+\hat{a}\right)\left(\hat{b}^{\dagger}+\hat{b}\right)-g\hat{a}^{\dagger}\hat{a}\left(\hat{b}^{\dagger}+\hat{b}\right) \nonumber \\
&+2\eta\beta \left(\hat{b}+\hat{b}^{\dagger}\right)^{3}+\frac{\eta}{4}\left(\hat{b}+\hat{b}^{\dagger}\right)^{4} +\hat{H}_{d,diss},
\end{align}
where, as in the main text, $\Delta_c = \Delta-2g\beta$ and $G=g\alpha$. The bath Hamiltonian is now given by~\cite{nonlinear1,Lemode_PRA_2015}:
\begin{align}
\hat{H}_{d,diss}=&\sum_{i}(\omega_{c,i}-\omega_l)\hat{A}_{i}^{\dagger}\hat{A}_{i}+\sum_{i}\omega_{m,i}\hat{B}_{i}^{\dagger}\hat{B}_{i} \nonumber \\ 
&+i\sqrt{\frac{\kappa}{2\pi\rho_{c}}}\sum_{i}\left(\hat{A}_{i}^{\dagger}\hat{a}-\hat{A}_{i}\hat{a}^\dag\right)+\hat{H}_{\gamma,int},
\end{align}
where, due to the transformation to the rotating frame, the cavity bath frequencies are shifted by $\omega_l$. Furthermore, fast oscillating terms in the interaction with the bath of the cavity can be safely omitted. 

The final Hamiltonian is obtained by applying the squeezing transformation $\hat{S}(r)$ given in the main text. The value of $r$ in Eq.~(\ref{r_squeezing}) is chosen to bring the mechanical Hamiltonian to a simple form:
\begin{align}
\hat{S}^\dag(r)\left[\Omega_m \hat{b}^\dag\hat{b} +6 \eta \beta^2 \left(\hat{b}^{\dag}+\hat{b} \right)^2\right]\hat{S}(r)=\widetilde{\Omega}_m \hat{b}^\dag\hat{b} ,
\end{align}
where $\widetilde{\Omega}_m = \Omega_m e^{2r}$. The other terms of Eq.~(\ref{Hd}) can be transformed by inspection, using
\begin{equation}\label{SxS}
\hat{S}^\dag(r) \left(\hat{b}+\hat{b}^\dag \right) \hat{S}(r) = \left(\hat{b}+\hat{b}^\dag \right) e^{-r}. 
\end{equation}
In particular, the dressed optomechanical coupling becomes $G_s = Ge^{-r}$, in agreement with Eq.~(\ref{eq:hamiltonian}). Applying Eq.~(\ref{SxS}) to the nonlinear terms, we immediately obtain Eq.~(\ref{Hnl}). Finally transforming Eq.~(\ref{Hint_gamma}) is equivalent to the rescaling of mechanical damping given in Eq.~(\ref{gamma squeezed}).

\section{General nonlinear interaction}\label{app:non_resonant}

In this Appendix, we discuss the effect of nonlinear terms neglected in Eq.~(\ref{eq:polariton}). They originate in a straightforward way from the nonlinear interaction of Eq.~(\ref{Hnl}), written in terms of polaritons. It is not difficult to see that the first line of Eq.~(\ref{Hnl}) generates all possible interactions of three polaritons. After normal ordering, they can be written as:
\begin{align}\label{non_res_terms_3}
\left(\sum_{i, j \geq k}\tilde{g}^{(1)}_{i,j,k} \hat{c}_{i}^\dag \hat{c}_{j} \hat{c}_{k} 
+\sum_{i\geq j\geq k}\tilde{g}^{(3)}_{i,j,k} \hat{c}_{i}^\dag\hat{c}_{j}^\dag \hat{c}_{k}^\dag\right)
 +{\rm H.c.},
\end{align}
where $i,j,k \in \pm$ (with $+ > -$). In particular, $\tilde{g}^{(1)}_{+,-,-}=\tilde{g}_{d}+\tilde{g}_{os}$ is the coupling of the resonant term included in Eq.~(\ref{eq:polariton}).  Effects of the non-resonant terms were included in the previous treatments without Duffing nonlinearity, where they can be seen to only give small shifts to the polariton energies~\cite{nonlinear1,florian2013,teuful2013}. Due to the formal analogy of the polariton Hamiltonian with and without Duffing nonlinearity (when expressed in polariton language), the case of our interest is completely analogous, except for the specific value of the coupling coefficients $g_{i,j,k}^{(1,3)}$. 

To estimate the  $g_{i,j,k}^{(1,3)}$, we restrict ourselves to the limit of large Duffing nonlinearity relevant to this work. Neglecting the contribution of the optomechanical nonlinearity [first term of Eq.~(\ref{Hnl})], we have:
\begin{equation}\label{gijk_estimate}
\tilde{g}^{(1,3)}_{i,j,k} \sim \eta\beta \varepsilon^{N_+},
\end{equation}
where $\varepsilon$ is a small hybridization parameter. Since $\hat{c}_- \simeq \hat{b}$ and $\hat{c}_+ \simeq \hat{a}$, the exponent $N_+=\delta_{+,i}+\delta_{+,j}+\delta_{+,k}$ simply counts the number of $+$ operators. As discussed in the main text, we have $\beta \sim \sqrt{\Omega_m/\eta} \gg 1$ while, from the transformation matrix $V$, one can infer $\varepsilon \sim \overline{\eta}^{-1/4}$.

As the non-resonant interaction terms couple states with energy difference of order $\Omega_m$, the corresponding energy shifts are of order $[\tilde{g}^{(1,3)}_{i,j,k}]^2/\Omega_m$. Then, we easily see from Eq.~(\ref{gijk_estimate}) that the largest effect is induced by the terms with $i=j=k=-$, where the small hybridization parameter $\varepsilon$ is absent ($N_+ =0 $). Using Eq.~(\ref{gijk_estimate}), we obtain that the shift in energy of the lower polariton is approximately given by:
\begin{equation}\label{dEminus_3pplaritons}
\delta E_- \sim  \frac{\left[\tilde{g}^{(1,3)}_{-,-,-}\right]^2}{\Omega_m} \sim \eta  ,
\end{equation}
which is usually negligible, since the largest achievable values of $\eta$ are typically many order of magnitudes smaller than $\kappa$ (see Sec.~\ref{exp_parameters}). More specifically, our main focus is on the self-energy of the photonic ($+$) polariton for which, as seen from Eq.~(\ref{eq:self2}), a change in $E_-$ induces a small shift in the associated OMIT dip. Considering the numerical case of Fig.~\ref{fig:desity}, the largest shift $\delta E_- \sim \eta $ is for the red curve, where $\eta= 10^{-4} \kappa$, thus is invisible on the horizontal frequency scale. Other non-resonant terms of Eq.~(\ref{non_res_terms_3}) give effects which are even smaller, due to the additional hybridization parameter. Explicitly, they are further suppressed by a small factor which is at most $[\tilde{g}^{(3)}_{+,-,-}]^2/[\tilde{g}^{(1,3)}_{-,-,-}]^2 \sim \varepsilon^2$.

In addition to the three-polariton terms discussed above, the second line of Eq.~(\ref{Hnl}) generates four-polariton interactions of the following form:
\begin{align}\label{non_res_terms_4}
&\sum_{i \geq j, k\geq l}\tilde{g}^{(2)}_{i,j,k,l} \hat{c}_{i}^\dag \hat{c}_{j}^\dag \hat{c}_{k} \hat{c}_{l}
+\left(\sum_{i, j \geq k \geq l}\tilde{g}^{(1)}_{i,j,k,l} \hat{c}_{i}^\dag \hat{c}_{j} \hat{c}_{k} \hat{c}_{l}
 \right. \nonumber \\
&\left. +\sum_{i\geq j\geq k \geq l}\tilde{g}^{(4)}_{i,j,k,l} \hat{c}_{i}^\dag\hat{c}_{j}^\dag \hat{c}_{k}^\dag \hat{c}_{l}^\dag +{\rm H.c.}\right),
\end{align}
where in the first term the real couplings satisfy $\tilde{g}^{(2)}_{i,j,k,l}=\tilde{g}^{(2)}_{k,l,i,j}$. In this case we estimate
\begin{equation}\label{gijkl_estimate}
\tilde{g}^{(1,2,4)}_{i,j,k,l} \sim \eta \varepsilon^{N_+},
\end{equation}
where $N_+=\delta_{+,i}+\delta_{+,j}+\delta_{+,k}+\delta_{+,l}$ is the number of $+$ operators. Among these interactions, the most notable are those of the form $\tilde{g}^{(2)}_{i,j,i,j}\hat{c}_{i}^\dag \hat{c}_{j}^\dag \hat{c}_{i} \hat{c}_{j}$, as they are diagonal in the polariton number basis. To lowest order, their effect is simply a shift of the polariton frequencies. The largest coupling is $\tilde{g}^{(2)}_{-,-,-,-} \sim \eta$, due to the absence of hybridization factors. Therefore, in the low-temperature limit (when zero-point fluctuations are taken into account), the shift is at most comparable to Eq.~(\ref{dEminus_3pplaritons}). Instead, at large temperatures:
\begin{equation}
\delta E_- \sim \eta \bar{n}_- .
\end{equation}
Despite the thermal enhancement factor, this shift is usually negligible since, as discussed in relation to Eq.~(\ref{dEminus_3pplaritons}), $\eta$ is typically very small. Furthermore, it does not affect the qualitative features of the OMIT peak, except for a slight shift in the position of the resonance. The effects of the other two terms, $\propto \hat{c}_{+}^\dag \hat{c}_{-}^\dag \hat{c}_{+} \hat{c}_{-}$ and $\propto \hat{c}_{+}^\dag \hat{c}_{+}^\dag \hat{c}_{+} \hat{c}_{+}$, are further suppressed by additional hybridization factors. Furthermore, at high temperature $\bar{n}_+ \ll \bar n_-$.

We finally turn to the off-diagonal terms of Eq.~(\ref{non_res_terms_4}), which are all non-resonant when the condition $E_+ = 2 E_-$ is satisfied. Here, compared to the three-wave mixing terms, the effective couplings are not enhanced by the large factor $\beta \gg 1$, due to the classical displacement of the mechanical mode. As a consequence, the same reasoning based on second-order perturbation theory yields much smaller energy shifts.

\section{Other multi-polariton processes}\label{appendix_multipolariton}

Besides the single-polariton considerations of the main text, relevant for the OMIT signal, the issue of multi-polariton effective interactions deserves further consideration. In the case without Duffing nonlinearity, the emergence at the resonance point of an effective Kerr interaction between phonon-like ($-$) polaritons has been discussed in Ref.~\cite{nonlinear1}. This effective interaction is mediated by the propagation of a photonic polariton, hence the appropriate energy denominator is $\kappa_+ \simeq \kappa$ which, in the case without Duffing nonlinearity, gives an interaction strength of order $g^2/\kappa$~\cite{nonlinear1}. For our case, assuming a large Duffing nonlinearity $\eta \gg g^2/\Omega_m$,  the effective Kerr interaction strength can be estimated as:
\begin{equation}
\frac{\tilde{g}_d^2}{\kappa} \sim g \frac{\sqrt{\eta \Omega_m}}{\kappa},
\end{equation}
and is indeed enhanced compared to $g^2/\kappa$, by a factor $\sqrt{\eta\Omega_m}/g = \overline \eta^{1/2}$. Notably, the above coupling can be larger than the strength of the equivalent quartic term induced directly by the Duffing interaction (which has coupling $g^{(2)}_{-,-,-,-}\simeq \eta$, see Appendix~\ref{app:non_resonant}). For example, using the parameters of Fig.~\ref{fig:ceff_exp}, taken from the experiment of Ref.~\cite{duffingex4}, one gets $\tilde{g}_d^2/\kappa \sim 10 \eta$. However, the effective interaction becomes negligible if larger values of $\eta$ can be realized (as desired). For example, with the much larger values of $\eta$ measured in Ref.~\cite{duffingex3}, the Duffing term would completely dominate the effective Kerr interaction. 

A different set of considerations applies to the alternative resonance condition allowed by the quartic form of the Duffing interaction:
\begin{equation}\label{three_phonons_resonsnce}
E_+ = 3 E_-.
\end{equation}
Although we will refrain from providing an accurate treatment, the typical scale of the expected effects can be easily obtained. First, we note that the value of $\beta$ at the resonant condition can be derived analogously to Eq.~(\ref{eq:beta_res2}) and is still given by $\beta \sim \sqrt{\Omega_m/\eta}$. Therefore, the polaritons are weakly hybridized and the hybridization parameter $\varepsilon$ of Eq.~(\ref{gijkl_estimate}) can be estimated as for the $E_+=2E_-$ resonance, i.e., $\varepsilon \sim \overline\eta^{-1/4}$. Finally, the resonant scattering process is now $\propto\hat{c}^\dag_+\hat{c}_-\hat{c}_-\hat{c}_- $ (and its Hermitian conjugate), with a coupling strength of order:
\begin{equation}
g^{(1)}_{+,-,-,-} \sim \eta \varepsilon.
\end{equation}
From this estimate we see that the effects on OMIT are greatly reduced with respect to the three-wave mixing term, as the interaction is not enhanced by the large value of $\beta$. More in detail, the effective damping of the $-$ polariton, involved in the virtual intermediate states, is still $\kappa_- \sim \kappa \varepsilon^2$, giving a second-order self-energy $\Sigma_+ \sim \eta^2/\kappa$ (much smaller than the previous result for the $E_+ = 2E_-$ resonance, $\Sigma_+ \sim \eta\Omega_m/\kappa$). The effective cooperativity now reads:
\begin{equation}
C_{eff} \sim \frac{\eta^2}{\kappa^2},
\end{equation}
instead of Eq.~(\ref{Ceff_estimate_Duffing}), indicating that the resonance condition in Eq.~(\ref{three_phonons_resonsnce}) is much less advantageous for observing strong nonlinear signatures in the OMIT signal.

As for the generation of an effective interaction between phonon-like polaritons, induced by the propagation of a virtual photon-like polariton, a similar mechanism takes place with $E_+ = 3 E_-$. The resulting coupling strength  $\sim [g^{(1)}_{+,-,-,-}]^2/\kappa$ is much weaker than the effective Kerr interaction discussed above (by a factor $\eta/\Omega_m$, due to the absence of $\beta^2$). It must be pointed out, however, that the effective interaction is qualitatively different, being of the type $(\hat c_-^{\dagger} \hat c_-)^3$. Therefore, its consequences might deserve further analysis.

\section{Higher-order expansion of $C_{eff}$\label{app:expansion}}
At large $\overline \eta$, we can expand  $C_{eff}$ as follows:
\begin{equation}\label{Ceff_general_expansion}
C_{eff} =\overline\eta \frac{g^2}{\kappa^2} \left(\sum_{n=1}^\infty   \frac{F_{n}(\Omega_m/\Delta)}{\overline\eta ^{(n-1)/2}} \right),
\end{equation}
where the first two terms give Eq.~(\ref{Ceff_next_order}):
\begin{align}
& F_1(x)= \frac{15}{2} x (1-4x^2), \\
& F_2(x)= \frac{1} {{36}\sqrt{6}} x\sqrt{1-4x^2}(1811-796x^2+3008 x^4), 
\end{align}
and the next term reads:
\begin{align}
F_3(x)= & \frac{x}{15552(1+8x^2)^2}(228549+3321544 x^2  \nonumber
\\ &+10522960 x^4-8897024 x^6+73398272 x^8 \nonumber \\ 
&+68943872 x^{10}-9109504 x^{12}).
\end{align}

\bibliography{references}

\end{document}